\documentclass[useAMS,usenatbib]{mn2e}

\usepackage[dvips]{graphicx}
\def\zsun{{\rm Z_\odot}}
\def\msun{{\rm M_\odot}}

\def\Omegat{{\Omega_{0,\rm tot}}}
\def\Omegab{{\Omega_{0,\rm b}}}
\def\Omegam{{\Omega_{0,\rm m}}}
\def\Omegal{{\Omega_{0,\rm \Lambda}}}

\renewcommand{\d}{\mathrm{d}}

\newcommand{\gtrsim}{\ga}
\newcommand{\lesssim}{\la}

\newcommand{\be}{\begin{equation}}
\newcommand{\ee}{\end{equation}}
\newcommand{\bea}{\begin{eqnarray}}
\newcommand{\eea}{\end{eqnarray}}
\newcommand{\bfig}{\begin{figure}}
\newcommand{\efig}{\end{figure}}
\newcommand{\bfigs}{\begin{figure*}}
\newcommand{\efigs}{\end{figure*}}
\newcommand{\bt}{\begin{table}}
\newcommand{\et}{\end{table}}
\renewcommand{\vec}[1]{ {\bf #1} }

\title[Radiative feedback and cosmic molecular gas]{
Radiative feedback and cosmic molecular gas: numerical method
}

\author[M.~Petkova \& U.~Maio]{
Margarita Petkova$^{1,2}$\thanks{E-mail: margarita.petkova@unibo.it},
Umberto~Maio$^{3}$\thanks{E-mail: umaio@mpe.mpg.de}\\
${}^1$University of Bologna, Department of Astronomy,
via Ranzani 1,  I-40127 Bologna, Italy\\
${}^2$Max-Planck-Institut f\"ur Astrophysik,
Karl-Schwarzschild-Stra{\ss}e 1,  D-85748 Garching b. M\"unchen,
Germany\\
${}^3$Max-Planck-Institut f\"ur extraterrestrische Physik,
Giessenbachstra{\ss}e 1,  D-85748 Garching b. M\"unchen, Germany 
}

\begin{document}

\date{(draft)}
\pagerange{\pageref{firstpage}--\pageref{lastpage}}\pubyear{0}
\maketitle
\label{firstpage}

\begin{abstract}
We present results from self-consistent numerical simulations of cosmic structure formation 
with a multi-frequency radiative transfer scheme and non-equilibrium molecular chemistry 
of 13 primordial species
(e$^-$, H, H$^+$, H$^-$, He, He$^+$, He$^{++}$, H$_2$, H$_2^+$, D, D$^+$, HD, HeH$^+$),
performed by using the simulation code {\small GADGET}.
We describe our implementation and show tests for ionized sphere expansion in a static 
and dynamic density field around a central radiative source, and for cosmological abundance 
evolution coupled with the cosmic microwave background radiation.
As a demonstrative application of radiative feedback on molecular gas, we run also
cosmological simulations of early structure formation in a $\sim 1\,\rm Mpc$ size box.
Our tests agree well with analytical and numerical expectations.
Consistently with other works, we find that ionization fronts from central sources can 
boost H$_2$ fractions in shock-compressed gas.
The tight dependence on H$_2$ lead to a corresponding boost of HD fractions, as well.
We see a strong
lowering of the the typical molecular abundances up to several orders of magnitudes
which partially hinders further gas collapse of pristine neutral gas, and
clearly suggests the need of re-ionized gas or metal cooling for the formation of the
following generation of structures.
\end{abstract}


%************************************************************************

\section{Introduction}\label{Sect:introduction}

%************************************************************************

The preset understanding of cosmic structure formation relies on the
observations of a Universe expanding at a rate of $H_0\simeq\rm
70\,km/s/Mpc$ and whose energy budget is largely dominated by a form
of unknown `dark' energy or cosmological constant, $\Lambda$, that
contributes $\sim70\%$ to the total cosmic energy content. 
The residual matter contribution is roughly $\sim 30\%$, but only a
very small fraction of $\sim 4\%$ consists of ordinary baryonic
matter, while the rest is unknown cold (i.e. non-relativistic) `dark
matter' (DM).
More precisely, recent determinations suggest
$\Omegam = 0.272$, $\Omegal = 0.728$, and $\Omegab = 0.044$
\cite[][]{Komatsu2011}.
In this framework, also called $\Lambda$CDM model, cosmological
structures can grow from gravitational instability \cite[][]{Jeans1902}
of primordial matter fluctuations, probably originated during the
primordial inflationary epoch.
These early perturbations represent the seeds which would develop into
present-day galaxies and stars \cite[][]{SS1953} by gas cooling and
condensation \cite[][]{Spitzer1962},
and they will affect the surrounding environment through a number of 
mechanical, chemical and radiative processes commonly known as 
feedback effects 
\cite[on this topic, see the extensive review by][]{CiardiFerrara2005}.
\\
Quantitatively speaking, linear perturbation analyses are usually
performed to study the initial phases of gravitational collapse, where
a Gaussian density distribution for the primordial matter fluctuations
is assumed.
The linear expansion of the continuity, Euler, and energy equations
can also be extended with higher-order corrections
\cite[e.g.][]{TseliakhovichHirata2010,Maio2011b,Stacy2011,Greif2011arXiv}
or non-Gaussian deviations
\cite[e.g.][]{Grinstein1986,Koyama1999,Komatsu2002,Grossi2007,Desjacques2009,MaioIannuzzi2011,Maio2011arXiv,MaioKhochfar2011arXiv},
but to study non-linear regimes and feedback mechanisms it is essential 
to perform numerical integrations and use N-body/hydro simulations.
Indeed, to capture early gas collapse it is needed not only to follow
gravity and hydrodynamics, but also its full chemistry evolution and
molecule formation.
Since in the cosmic medium hydrogen (H) is the most abundant species
with a cosmological mass fraction of $X_{\rm H}\simeq 0.76 $
(corresponding to $\sim 0.93$ in number fraction), its contribution in
gas cooling is likely to play a very relevant role, together with
helium (He).
However, H and He collisional processes are able to cool the medium
to $\sim 10^4\,\rm K$ via resonant line transitions, but they are
not capable to bring the gas temperature further down.
At such low temperatures, thermal collisions are not able to
excite the electrons to higher levels due to the large energy gaps
(from a few to some tens of eV) of H and He atomic configurations.
\cite{SaZi1967} proposed that gas cooling and fragmentation could be
continued below $\sim 10^4\,\rm K$ by H$_2$ cooling, down to $\sim
10^2-10^3\,\rm K$.
Later, \cite{LeppShull1984} suggested that the existence of primordial
deuterium (D) could determine HD formation and consequent cooling even
below $\sim 10^2\,\rm K$, down to several $\sim 10\,\rm K$. 
During the last decades, these problems have been tackled by collecting
full reaction networks 
\cite[][]{ShapiroKang1987,Puy_et_al_1993,GP98,HuiGnedin1997,Abel_et_al1997,UeharaInutsuka2000,NakamuraUmemura2002,Omukai2005,GloverAbel2008,Omukai2010,Glover2010}
and by running high-resolution chemistry cosmological
simulations, both in the standard $\Lambda$CDM model
\cite[e.g.][]{Abel2002,Bromm2002,Yoshida2003,Yoshida2007} and in dark-energy 
cosmological models \cite[][]{Maio2006}.
Further on, the effects of metal cooling have been investigated in
numerical simulations by joining molecular chemistry evolution with
metal pollution and low-temperature fine structure transitions from,
e.g., C, O, Si, Fe \cite[e.g.][]{Maio2007}.
These studies clearly show the strong implications of metals on the
cosmological chemical evolution (chemical feedback) and how the 
rapidity of early enrichment from first star formation episodes
\cite[see also][]{Tornatore2007} overcomes gas molecular cooling
\cite[][]{Maio2010,Maio2011a}, marking  the transition from the
primordial, pristine star formation regime - i.e. the so-called
population III (popIII) regime - to the more standard population II-I
(popII-I) regime \cite[][]{Maio2010}.
This transition is often parameterized in terms of a minimum critical
metallicity $Z_{crit}\sim 10^{-4}\zsun$, but given our ignorance about
early dust production and detailed atomic and molecular data,
$Z_{crit}$ has large uncertainties: its expected value  varies between
$\sim 10^{-6}\zsun$ \cite[][]{Schneider2003} -- $ 10^{-5}\zsun$
\cite[][]{Omukai2010} and $\sim 10^{-3}\zsun$ \cite[][]{Bromm_Loeb_2003}.
  There are also recent arguments on the possible existence
  of a critical dust-to-gas ratio \cite[][]{Schneider2011arXiv}.\\
Once the first stars are formed, they shine and emit radiation.
This can have relevant impacts (radiative feedback) 
\cite[][]{Efstathiou1992} on the surrounding medium and on the following 
star formation history
\cite[e.g.][]{ciardi2000a,Ciardi2001,Whalen2004,Iliev2005,Alvarez2006,SusaUmemura2006,Iliev2006,Iliev2009,Whalen2010,Mellema2006,Susa2009,Gnedin_et_al_2009,Petkova2011,Paardekooper2011,Wolcott-Green2011tmp}.
Therefore, appropriate radiative transfer (RT) calculations coupled
with hydrodynamics and chemistry must be performed in order to consistently
explore the repercussions on molecule destruction or enhancement,
and hence on structure formation in the early Universe.
There have been studies on how the radiation feedback affects larger
halos such as clusters and galaxies 
\cite[e.g.][]{Shapiro2004,Yoshida2007,Croft2008},
and particular interest has arisen from popIII stars
\cite[][]{WiseAbel2008,Hasegawa2009,Smith2009,Whalen2010},
whose large luminosities
\cite[which suggest them as progenitors of early gamma-ray bursts, e.g.][]{Campisi2011a,deSouza2011,Campisi2011b}
not only ionize, but also expel gas from the pristine mini-halos they sit in.
For example, Lyman-Werner photons ($11.2 - 13.6$~eV) at early times could 
eradicate H$_2$ from halos, delaying or completely impeding the collapse of
molecular gas
\cite[e.g.][]{Susa2007,Wise2007,Johnson2007,Ahn2009,TrentiStiavelli2009,Whalen2010}.
Non the less, several issues in this respect are still unsolved, like 
the role of radiative feedback on popIII star formation, its efficiency
in halting or boosting gas cooling in primordial environment, its interplay with
mechanical and chemical feedback, its effects on the dynamical and
thermodynamical state of the cosmic gas, or its connections to the 
set-up of a turbulent medium dominated by hydro-instabilities 
\cite[e.g.][]{Maio2011b}.
\\
First semi-analytical works 
\cite[][]{Haiman1997,Haiman1997erratum,Haiman1999}
expected molecular hydrogen to be universally destroyed
by UV stellar radiation in the LW band.
\\
Subsequent, one-dimensional, numerical studies of small- or intermediate-size 
boxes ($<1\,\rm Mpc$ a side), including RT coupled with hydrodynamics,
demonstrated that these predictions were overestimating 
molecular destruction, and, despite the stellar UV radiation, 
H$_2$ could be rapidly reformed in the shock compressed gas
of the ionization fronts (I-fronts) of HII regions
\cite[][showed that via a Softened Lagrangian Hydrodynamics Particle-Particle-Particle-Mesh, SLH-P$^3$M, implementation]{Ricotti2001,Ricotti2002a,Ricotti2002b}.
This was the first suggestion for ``positive'' feedback 
exerted by I-fronts on primordial structure formation.
\\
Similar calculations applied to individual haloes nearby primordial stellar 
sources \cite[e.g.][]{Shapiro2004,Iliev2005,Hasegawa2009} 
showed that UV radiation from popIII stars in crowded star forming 
regions could photoevaporate small ($\sim 10^7\,\rm \msun$) haloes 
in $\gtrsim 100\,\rm Myr$.
However, more recent studies
\cite[e.g.][]{SusaUmemura2006,Susa2009,Hasegawa2009,Whalen2008,WhalenNorman2008,Whalen2010}
found that halo photoevaporation due to first stars strongly depends 
on the features of the stellar sources and on the hydrodynamical 
properties of the collapsing cloud.
Thus, radiative feedback was found to be able to trigger gas
cooling by catalyzing H$_2$ formation.
\\
In order to overcome numerical issues in larger cosmological 
simulations ($\gtrsim 1\,\rm Mpc$ a side), further
studies adopted simple analytical prescriptions for a
mean, {\it uniform}, UV background, assumed to be established after 
the early onset of star formation
\cite[e.g.][via a grid Adaptive Mesh Refinement, AMR, implementation]{Machacek2001,Machacek2003,Mesinger2006,Mesinger2009}.
In the same spirit, detailed, AMR, popIII, star formation calculations 
in {\it uniform} LW backgrounds 
\cite[][]{OSheaNorman2008,WiseAbel2008,Shang2010}
confirmed that the LW background is much less destructive to 
new popIII stars than originally supposed, and, even by assuming 
extremely high background values of 
$J \ge 10^{-21}\,\rm erg\, s^{-1}\, cm^{-2}\,Hz^{-1}\, sr^{-1}$
($\equiv J_{21}$) star formation is just delayed, not shut down.
\\
This essential review shortly shows how debated the issues related to 
radiative feedback are and highlights the numerous unknowns when 
dealing with RT and molecular chemistry.
In the present work, we aim at contributing to the scientific 
discussions and works already existing in literature by implementing 
three-dimensional RT schemes fully coupled with hydrodynamics, and
non-equilibrium chemistry.
At the end of the paper, together with chemistry and radiative transfer,
we will also consider additional feedback mechanisms which are of 
relevant interest for cosmological structure formation, namely, SN feedback 
and wind feedback.
This represents a step further with respect to previous implementations,
since they usually focus on individual feedback processes.
We perform the implementation within the widely used and well-tested
numerical N-body/SPH code {\small GADGET} \cite[][]{Springel2005},
in its most recent and updated version.
In this way, we will be able to self-consistently study the effects
of RT on chemical abundances, and to pinpoint the basic consequences 
for the cosmological evolution of the structures in the Universe.
\\
In Sect.~\ref{Sect:implementations} we describe the implementations of
radiation (Sect.~\ref{Sect:radtrans}) and cosmic chemistry evolution
(Sect.~\ref{Sect:chemistry}).
In Sect.~\ref{Sect:simulations}, we test our implementation by
performing analyses of the Str{\"o}mgren sphere problem
(Sect.~\ref{Sect:ss}) in a static (Sect.~\ref{sec:static}) and
dynamic (Sect.~\ref{sec:dyn}) density field, and chemical abundance evolution
(Sect.~\ref{Sect:abundances}).
We then apply our method to cosmological structure formation simulations
(Sect.~\ref{Sect:cosmo}).
We summarize, discuss and conclude in Sect.~\ref{Sect:discussion}.

%************************************************************************

\section{Implementations}\label{Sect:implementations}

%************************************************************************

We use the parallel tree N-body/SPH (smoothed particle hydrodynamics)
code {\small GADGET3}, an extended version of the publicly available 
code {\small GADGET2} \cite[][]{Springel2005}, and modify it  in order to 
couple chemistry evolution and RT. 
In the following paragraphs, we give the details about the
implementations of the RT (Sect.~\ref{Sect:radtrans})
and chemistry (Sect.~\ref{Sect:chemistry}) parts, and, in the next
section, we will show results from our test runs. 

%************************************************************************

\subsection{Radiative transfer}\label{Sect:radtrans}
To follow the propagation of radiation we use the implementation of
RT in {\small GADGET3} \cite[][]{Petkova2009} and
expand it to a multi-frequency scheme. The original implementation is
based on a moment method, where the closure relation is the optically
thin variable Eddington tensor, suggested by \cite{GA2001}.
To follow the transport of radiation, we solve the equation of
anisotropic diffusion for the photon number 
density per frequency $n_\gamma(\nu)$:  
\be
\frac{\partial n_\gamma(\nu)}{\partial t}= c \frac{\partial}{\partial
x_j}\left( \frac{1}{\kappa(\nu)}\frac{\partial n_\gamma(\nu) h^{ij}}{\partial
x_i} \right) - c\,\kappa(\nu)\, n_\gamma(\nu) + s_\gamma(\nu) ,
\label{eqn:rt}
\ee
where
$t$ is time,
$x_i$ and $x_j$ are the coordinate components,
$c$ is the speed of light, 
$\kappa(\nu)$ is the absorption coefficient, 
$h^{ij}$ are the components of the Eddington tensor,
$s_\gamma(\nu)$ is the source function, and
the Einstein summation convention is adopted for all exponents $i$ and
$j$.

The Eddington tensor is obtained by summing up the contributions from
the sources of ionizing photons and its components are given by
\be h^{ij} = \frac{P^{ij}}{{\rm Tr}(P^{ij})}\, ,\ee where
\be
P^{ij}(\vec{x}) = \int {\rm d}^3 x' \rho_{\ast}(\vec{x}')
\frac{(\vec{x}-\vec{x}')_i(\vec{x}-\vec{x}')_j}
     {(\vec{x}-\vec{x}')^4}\, 
\ee
is the radiation pressure tensor and $\rho_{\ast}$ is the stellar density. The tensor is computed via
a tree and effectively removes the dependence of the scheme on the
number of ionizing sources -- an advantage in cosmological and galaxy
formation simulations.

The source term is treated in a {\it Strang-split} fashion, where photons are 
first ``injected'' into the medium surrounding the sources, and are
then ``diffused'' via equation~(\ref{eqn:rt}). Solving the equation
for all particles in the simulation reduces to a linear system of
equations, which we solve implicitly by using a Conjugate-Gradient scheme, that
ensures robustness and stability even for large timesteps.

For our multi-frequency extension we need to transform the photon
number density to ionizing intensity and vice versa. 
Since the photoheating and photoionization rates 
in equations~(\ref{rad_rates})
(discussed more in detail in Sect.~\ref{Sect:chemistry}) are obtained by 
integrating the intensity over frequency, we use a 
single photon number density in each frequency bin.

The photon number density per frequency is derived from the ionizing intensity,
$I(\nu)$, as
\be
\label{n_gamma1}
n_{\gamma}(\nu) = \frac{1}{c} \frac{ 4\pi I(\nu)}{h_p \nu} \, ,
\ee
where $4\pi$ is the full solid angle and $h_p$ is the Planck constant.
For any particle and at any timestep, the RT
equation~(\ref{eqn:rt}) is then solved for each frequency bin.

\bfig
\centering
\includegraphics[width=0.45\textwidth]{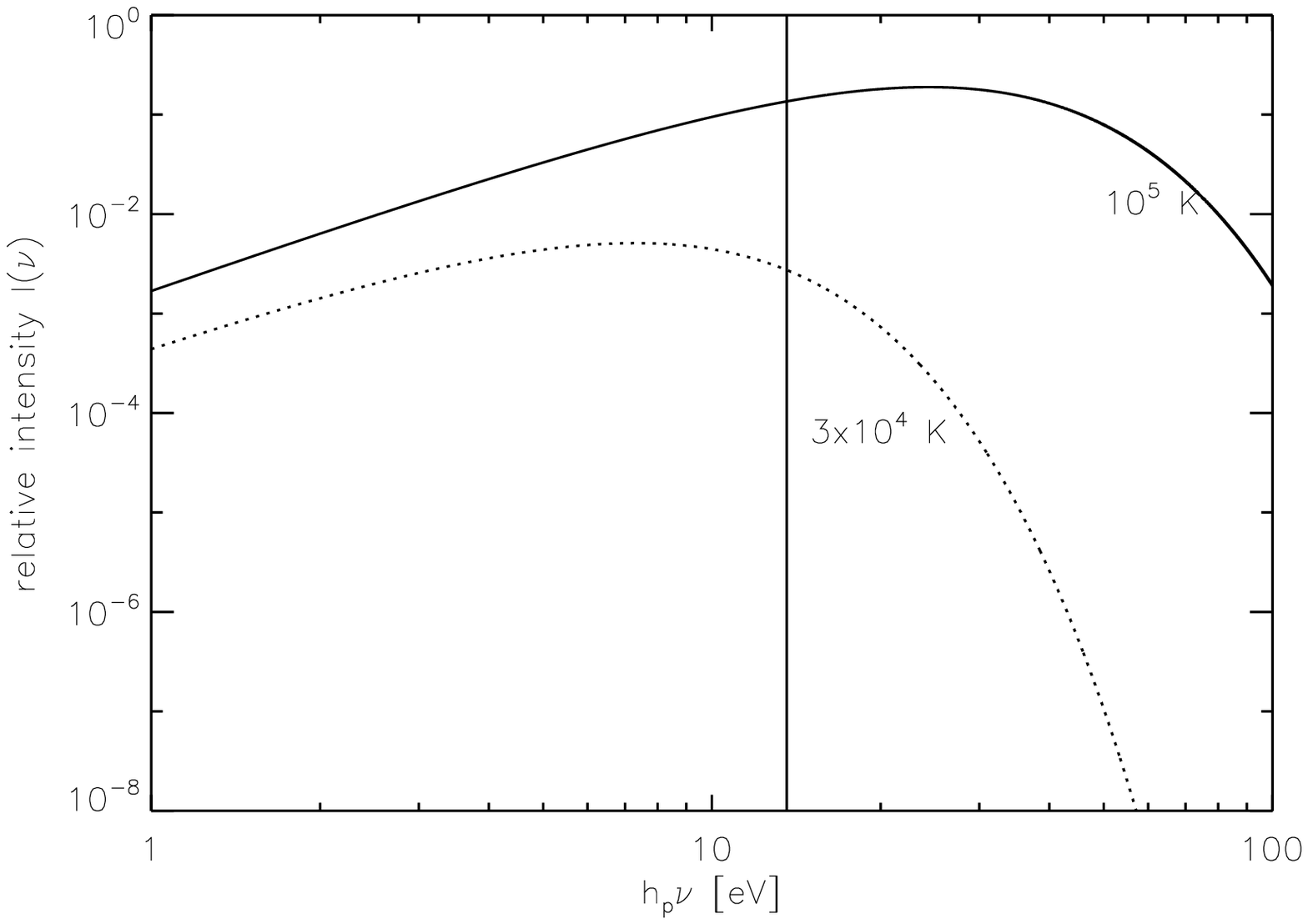}
\caption[]{\small Intensity of a black-body spectrum as a function of
  photon energy in eV for two different
  effective temperatures - $3 \times 10^4 \, \rm K$ and $10^5 \, \rm
  K$. The vertical line marks the end of the Lyman-Werner band at
  $13.6 \, \rm eV$.
 }
\label{fig:BB}
\efig

The intensity of the radiative source, $I(\nu)$, depends on the particular 
problems treated.
For stellar sources, a common, simple and suitable approximation (which we
will also adopt in the following) is a black-body spectrum 
(see Fig.~\ref{fig:BB}), 
with effective temperature dependent on the assumed stellar population.
This treatment allows us to take into account contributions from a wide 
frequency range, even below the H-ionization energy of 13.6~eV.
Indeed, as evident from the plot in Fig.~\ref{fig:BB} the low-energy tail
of the black-body spectrum contributes to the Lyman-Werner band and that
will definitely affect the molecular evolution of the gas (see later).

We note that dealing with radiation below $\sim$~13.6~eV is
a very debated and complicated problem
\cite[e.g.][]{ThoulWeinberg1996,Haiman1997,Haiman1997_Erratum,OmukaiNishi1999,Machacek2001,Kitayama2001,Ricotti2002a,Ricotti2002b,Mackey2003,Shapiro2004,Dijkstra2004,SusaUmemura2004,Stacy2011arXiv},
because of the many lines involved (76) in the LW band.
Full, detailed modeling of RT in such regimes is beyond the aims of this 
work, and, at some levels, it might be superfluous \cite[e.g.][]{Ricotti2001},
since the ionizing flux
at high redshift ($z$) is dominated by radiation from neighboring 
haloes \cite[e.g.][]{Ciardi2000b}, rather than from the soft-UV background
in the LW band.
Thus, H$_2$ photodissociation is simply accounted for by using
\cite[e.g.][]{Abel_et_al1997,Ricotti2001,Yoshida2003,Ahn2007,Maio2007,WhalenNorman2008} 
the radiative rate obtained from integration of the source intensity
over the LW range, $11.2-13.6$~eV -- see eq.~(\ref{rad_rates}), 
in the next Sect.~\ref{Sect:chemistry} -- shielded according to 
\cite{DraineBertoldi1996}.

%************************************************************************

\subsection{Chemistry}\label{Sect:chemistry}

\begin{table}
  \begin{center}
    \caption{\small Ionization energies in [eV]}
    \label{tab:Eion}
    \begin{tabular}{cccccccc}
      \hline
      H    & H$^-$& He  & He$^+$ & D   & H$_2$  & HD & HeH$^+$ \\
       & & & & & & & \\
      13.6 & 0.7 & 24.6& 54.4   & 14.9 & 15.4 & 15.4 &  44.5\\
      \hline
    \end{tabular}
  \end{center}
  {}
\end{table}

To couple radiation with chemistry, we include non-equilibrium
reactions for H, He and molecule evolution (whose ionization and 
dissociation energies are quoted in Tab.~\ref{tab:Eion}),
by following a chemical network (see Tab.~\ref{tab:reactions}) of
several species: 
e$^-$, H, H$^+$, H$^-$, He, He$^+$, He$^{++}$, H$_2$, H$_2^+$, D, 
D$^+$, HD, HeH$^+$.
Besides some updates in the rates and in the reaction network,
the implementation used is the same as the
one in \cite{Maio2007} and \cite{Maio2010}
\cite[and based on][]{Abel_et_al1997,GP98,Yoshida2003}.
Since the main coolants at early times are H-derived molecules, H$_2$
\cite[e.g.][]{SaZi1967} and HD \cite[e.g.][]{LeppShull1984}, the
inclusion of a large network is crucial to correctly resolve the
hydrodynamics and the fragmentation processes of high-redshift gas, as
well demonstrated by e.g. \cite{Abel_et_al1997},
\cite{Yoshida2003,Yoshida2006},
\cite{Maio2006,Maio2007,Maio2009,Maio2010, Maio2011a},
\cite{Maio2011b,MaioIannuzzi2011}.

To take into account chemical evolution, at each timestep and for each
species $i$, the time variation of its number density $n_i$ is
computed, for collisional and photoionization/photodissociation
events, via 
\begin{equation}\label{noneq_eq}
\frac{\d n_i}{\d t}= \sum_p\sum_q k_{pq,i} n_p n_q -  \sum_l k_{li} n_l n_i - k_{\gamma i}n_i,
\end{equation}
where $k_{pq,i}$ is the rate of creation of the species $i$ from
species $p$ and $q$, $k_{li}$ is the destruction rate of the species
$i$ from collisions with species $l$, and $k_{\gamma i}$ is the
photoionization or photodissociation rate of species $i$ due to
radiation (see Tab.~\ref{tab:Eion}).
\\
The collisional rates are given by
\begin{equation}
\label{collisional_rates}
k_{pq,i} = \int u \sigma_{pq,i}(u) f(u) \d^3 u,
\end{equation}
(and the analogous for $k_{li}$), with $u$ relative velocity of
particles $p$ and $q$, $\sigma_{pq,i}(u)$ interaction cross-section,
and $f(u)$ Maxwellian velocity distribution function.
More precisely, we do not compute the integrals on the fly, but
instead we interpolate pre-computed tables (whose references are 
listed in Tab.~\ref{tab:reactions}) in order to speed-up the code.
These rates are temperature dependent and are expressed in units of
volume per time, i.e. $\rm [cm^3\, s^{-1}]$ in the cgs system. 
As in the left-hand side, the first and second term in the right-hand
side of equation (\ref{noneq_eq}) have dimensions of a number density
per unit time, $\rm [cm^{-3}\,s^{-1}]$ in the cgs system. 
\\
Consistently with Sect.~\ref{Sect:radtrans}, when the species $i$
interacts with radiation ($\gamma$) -- see Table~\ref{tab:reactions} --
and gets photo-ionized (like for H, D, H$^-$, He, He$^+$) or
photo-dissociated (like for H$_2$, H$_2^+$, HD, HeH$^+$), the corresponding
radiative rate $k_{\gamma i}$ can be written as
\begin{equation}
\label{rad_rates}
k_{\gamma i} = \int \frac{4 \pi I(\nu)}{h_p \nu} \sigma_{\gamma i}(\nu)\d\nu
= \int c \, \sigma_{\gamma i}(\nu) n_\gamma (\nu) \d\nu ,
\end{equation}
where the $4\pi$ stands for isotropic radiation, $I(\nu)$ is the source
intensity as a function of frequency $\nu$, $\sigma_{\gamma i}(\nu)$
is the cross-section for the given process, $h_p$ is the Planck
constant, $c$ the speed of light, and $n_\gamma (\nu)$ the
photon number density per frequency. 
In the final equality we made use of eq.~(\ref{n_gamma1}),
to formally get the rate expression similar to
eq.~(\ref{collisional_rates}).
The radiative rates are probabilities per unit time, and are given in
$\rm [s^{-1}]$ in the cgs system. 
So, the number of radiative interactions per unit time and volume between 
photons and $n_i$ particles is $k_{\gamma i} n_i$.
The latter quantity is added on the right-hand side of 
equation~(\ref{noneq_eq}) 
when photon interactions are taken in consideration,
and consistently with the other terms in the equation, this is also
given in units of number density per time, i.e. $\rm [cm^{-3}\,
 s^{-1}]$ in the cgs system.
\\
When talking about radiative interactions, one has to consider
that, while ionization energies (see Tab.~\ref{tab:Eion})
are uniquely defined, molecular dissociation energies might depend on 
the particular radiative process considered, and thus
all the various channels must be taken into account.
For example (see Tab.~\ref{tab:reactions}), LW radiation above the 
energy threshold of 11.2~eV can dissociate H$_2$ in 2H, but harder 
photons with energies larger than 15.42~eV would simply ionize the 
residual H$_2$ into H$_2^+$ + $e^-$.
Also for H$_2^+$ there are two possible branches: 
one above an energy threshold of 2.65~eV and below 21~eV 
($\rm H_2^+ + \gamma \rightarrow H^+ + H$), and a second one between
30~eV and 70~eV ($\rm H_2^+ + \gamma \rightarrow 2 H^+ + e^-$).
For the HeH$^+$ radiative interaction considered in our network
the energy threshold is about 1.7~eV.
\\

The set of differential equation~(\ref{noneq_eq}) is integrated via
simple linearization, so, given the timestep $\Delta t$, at each time
$t$ the temporal variation of the number fraction of species $i$ can
be written as 
\begin{equation}\label{discrete}
\frac{n_i^{t+\Delta t} - n_i^{t}}{\Delta t} = C_i^{t+\Delta t} -
D_i^{t+\Delta t} n_i^{t+\Delta t}, 
\end{equation}
where we have introduced the creation coefficient for the species $i$
\begin{equation}
\label{C}
C_i = \sum_p \sum_q  k_{pq,i} n_p n_q,
\end{equation}
in $\rm [cm^{-3} s^{-1}]$, and the destruction coefficient
\begin{equation}
\label{D}
D_i = \sum_l k_{li} n_l + k_{\gamma i},
\end{equation}
in $\rm [s^{-1}]$.
The contribution from photoionization or photodissociation is
accounted for by adding, in equation~(\ref{D}), the $k_{\gamma i}$
rates. 
The number density is updated from equation~(\ref{discrete}):
\begin{equation}
n_i^{t+\Delta t} = \frac{C_i^{t+\Delta t} \Delta t + n_i^t}{1 + D_i^{t+\Delta t}\Delta t }.
\end{equation}
We apply this treatment to all the chemical species included, with the
coefficients for each reaction in the network quoted in
Table~\ref{tab:reactions}.
Gas cooling or heating is computed from H and He collisional
excitations \cite[][]{Black1981,Cen1992}, ionizations
\cite[][]{Abel_et_al1997}, recombinations \cite[][]{HG1997}, H$_2$ and
H$_2^+$ emissions \cite[][]{GP98}, HD emissions
\cite[][]{HDcoolingfct}, Compton effect, and Bremsstrahlung
\cite[][]{Black1981}.
\\
The timestepping is limited by the cooling time,
\begin{equation}
\label{tcool}
t_{cool} = \left | \frac{E}{\dot E} \right |,
\end{equation}
and by the electron recombination time,
\begin{equation}
\label{telec}
t_e = \left | \frac{n_e}{\dot n_e} \right |,
\end{equation}
where $E$ and $n_e$ are the energy and the electron number fraction of
each particle, and $\dot E$ and $\dot n_e$ the corresponding time
variations.
For accuracy reasons in the abundance determinations, the chemical 
subcycles are done over 1/10 of $\rm min(t_{cool}, t_e)$
\cite[as described by e.g.][]{Anninos1997,Abel_et_al1997,Yoshida2003,Maio2007}.
The additional constrain given by $t_e$ is useful mostly when $n_e$
changes very steeply, like behind the shock fronts, or at $\sim
10^4\,\rm K$, below which hydrogen recombines very efficiently and
above which hydrogen gets ionized very fast. 
Further details can be found in \cite{Maio2007} and references
therein. 

\begin{table}
\begin{center}
\caption{\small Reaction network}
\label{tab:reactions}
\begin{tabular}{lr}
\hline
\hline
Reactions & References for the rate coefficients\\
\hline
 	H    + e$^-$   $\rightarrow$ H$^{+}$  + 2e$^-$ & A97 / Y06 / M07\\
	H$^+$   + e$^-$  $\rightarrow$ H     + $\gamma$    & A97 / Y06 / M07\\
        H + $\gamma$ $\rightarrow$  H$^+$ + e$^-$  & A97 / Y06 / M07 \\
	He   + e$^-$   $\rightarrow$ He$^+$  + 2e$^-$    & A97 / Y06 / M07\\
	He$^+$  + e$^-$   $\rightarrow$ He   + $\gamma$     & A97 / Y06 / M07\\
        He + $\gamma$ $\rightarrow$  He$^{+}$ + e$^-$   & A97 / Y06 / M07 \\
	He$^+$  + e$^-$   $\rightarrow$ He$^{++}$ + 2e$^-$    & A97 / Y06 / M07\\
	He$^{++}$ + e$^-$   $\rightarrow$ He$^+$  + $\gamma$ & A97 / Y06 / M07\\
        He$^+$ + $\gamma$ $\rightarrow$  He$^{++}$ + e$^-$  & A97 / Y06 / M07 \\
	H    + e$^-$   $\rightarrow$ H$^-$    + $\gamma$     & GP98 / Y06 / M07\\
        H$^-$ + $\gamma$ $\rightarrow$  H + e$^-$  & A97 / Y06 / M07 \\
	H$^-$    + H  $\rightarrow$ H$_2$  + e$^-$       & GP98 / Y06 / M07\\
        H    + H$^+$ $\rightarrow$ H$_2$$^+$  + $\gamma$ & GP98 / Y06 / M07\\
        H$_2^+$ + $\gamma$ $\rightarrow$ 2 H$^+$ + e$^-$ & A97 / Y06 / M07\\
        H$_2^+$ + $\gamma$ $\rightarrow$  H + H$^+$  & A97 / Y06 / M07\\
        H$_2$$^+$  + H  $\rightarrow$ H$_2$  + H$^+$     & A97 / Y06 / M07\\
	H$_2$   + H   $\rightarrow$ 3H            & A97 / M07\\
	H$_2$   + H$^+$ $\rightarrow$ H$_2$$^+$  + H     & S04 / Y06 / M07\\
  	H$_2$   + e$^-$   $\rightarrow$ 2H   + e$^-$ & ST99 / GB03 / Y06 / M07\\
      	H$^-$    + e$^-$   $\rightarrow$ H    + 2e$^-$   &A97 / Y06 / M07\\
       	H$^-$    + H   $\rightarrow$ 2H    + e$^-$       & A97 / Y06 / M07\\
       	H$^-$    + H$^+$ $\rightarrow$ 2H                &P71 / GP98 / Y06 / M07\\
       	H$^-$    + H$^+$ $\rightarrow$ H$_2$$^+$  + e$^-$& SK87 / Y06 / M07\\
        H$_2$$^+$  + e$^-$   $\rightarrow$ 2H            &GP98 / Y06 / M07\\
        H$_2$$^+$  + H$^-$  $\rightarrow$ H    + H$_2$   &A97 / GP98 / Y06 / M07\\
        H$_2$ + $\gamma$ $\rightarrow$  H$_2^+$ +  e$^-$  & A97 / Y06 / M07 \\
        H$_2$ + $\gamma$ $\rightarrow$ 2 H & A97 / R01 / Y03 / M07 \\
        D    + H$_2$   $\rightarrow$   HD   + H     & WS02 / M07\\
        D$^+$  + H$_2$   $\rightarrow$   HD   + H$^+$  & WS02 / M07\\
        HD   + H   $\rightarrow$   D   + H$_2$         & SLP98 / M07\\
        HD   + H$^+$  $\rightarrow$   D$^+$  + H$_2$   & SLP98 / M07\\
        H$^+$  + D   $\rightarrow$   H    + D$^+$   & S02 / M07\\
        H    + D$^+$  $\rightarrow$   H$^+$  + D    & S02 / M07\\
        D$^+$  + e$^-$  $\rightarrow$   D + $\gamma$    & GP98 \\
        D  + $\gamma$  $\rightarrow$   D$^+$  + e$^-$  & GP98 \\
        He    + H$^+$  $\rightarrow$   HeH$^+$  + $\gamma$    & RD82/ GP98 / M07\\
        HeH$^+$    + H $\rightarrow$   He  + H$_2^+$    & KAH79 / GP98 / M07\\
        HeH$^+$    + $\gamma$ $\rightarrow$   He  + H$^+$    & RD82 / GP98 / M07\\
\hline
\hline
\end{tabular}
\end{center}
Notes:
$\gamma$ stands for photons;
P71~=~\cite{Peterson1971};
KAH79~=~\cite{KAH1979};
RD82~=~\cite{RD1982};
SK87~=~\cite{SK1987};
A97~=~\cite{Abel_et_al1997};
GP98~=~\cite{GP98};
SLP98~=~\cite{SLD_1998};
ST99~=~\cite{ST99};
R01~=~\cite{Ricotti2001};
WS02~=~\cite{Wang_Stancil_2002};
S02~=~\cite{Savin_2002};
GB03~=~\cite{GB03};
Y03~=~\cite{Yoshida2003};
S04~=~\cite{Savin_et_al2004};
Y06~=~\cite{Yoshida2006_astroph};
M07~=~\cite{Maio2007}.
\end{table}

%************************************************************************

\section{Test simulations}\label{Sect:simulations}

%************************************************************************

In order to test the implementation we perform numerical simulations
under different conditions.
First, we numerically solve an expanding ionized sphere problem 
(Sect.~\ref{Sect:ss}) by fully including both the RT and chemistry treatments.
Then we show the cosmic evolution of the different chemical species,
coupled with the radiative gas emissions (Sect.~\ref{Sect:abundances}).
Finally, we will perform cosmological simulations of early structure formation
(Sect.~\ref{Sect:cosmo}) to check the effects of radiative feedback on gas
cooling and collapse.

For all simulations we use a set of 284 frequencies, covering
the range from $0.7$~eV to $100$~eV.
The density of frequency bins around the peaks of the photoionization
and photodissociation
cross-sections is increased - i.e. there are more frequency bins in the 
spectral regions of interest.
In this way we can ensure that photons are both traced and absorbed
properly and no spectrum-averaging mistakes are made.

%************************************************************************

\subsection{Ionized sphere expansion}\label{Sect:ss}

\subsubsection{Ionized sphere expansion in a static density
  field}\label{sec:static}

The expansion of an ionization front in a static, homogeneous and
isothermal gas is the only problem in radiation hydrodynamics that has
a known analytical solution and is therefore the most widely
used test for RT codes \cite[e.g.][]{Iliev2006,Iliev2009}.
For such a set-up, the ionized bubble around
the ionizing source reaches a final steady radius, called the
Str{\"o}mgren radius, where absorptions and recombinations are
balanced along the line of sight.
For an H-only gas, the Str{\"o}mgren radius is analytically given by
\be
r_{\rm  S} = \left(\frac{3\dot{N}_\gamma}{4\pi\alpha_{\rm B} n_{\rm H}^2}\right)^{1/3} ,
\label{rs1}
\ee
with $\dot{N}_\gamma$ -- the luminosity of the source in photons per second;
$\alpha_{\rm B}$ -- the case-B recombination coefficient;
and $n_{\rm H}$ -- the hydrogen number density.
The case-B recombination coefficient assumes the so called 
'on-the-spot' approximation, where photons from recombinations
to lower energy levels are immediately absorbed in the vicinity 
of their emission \cite[e.g.][]{Spitzer1978}.
If we approximate the ionization front (I-front) as infinitely thin,
i.e.~it features a discontinuity in the ionization fraction,
the temporal expansion of the Str\"omgren radius can be solved 
analytically in closed form, with the I-front radius $r_{\rm I}$  given by
\be
\label{rI}
r_{\rm I} = r_{\rm S}[1-\exp(-t/t_{\rm  rec})]^{1/3},
\ee
where
\be
t_{\rm rec} = \frac{1}{n_{\rm H}\alpha_{\rm B}}
\ee
is the recombination time.

In our first test we perform an ionized sphere expansion, but we allow
the temperature of the gas to vary in order to test the coupling between 
the RT and the full non-equilibrium chemistry treatment.
As a reference, we compare to the analytical case with constant temperature.
We follow the expansion of an ionized sphere around a source that
emits $\dot N_\gamma = 5 \times 10^{48} \, \rm photons \, s^{-1}$.
The shape of the source spectrum corresponds to a $3 \times 10^4\, \rm K$
black body.
The surrounding gas density is $\rho = 1.7 \times 10^{-27} \, \rm g \,
cm^{-3}$ ($\sim 10^{-3}\,\rm cm^{-3}$) and is sampled by $16^3$,
$32^3$, and $64^3$ gas 
particles\footnote{
For more resolution studies, see \cite{Petkova2009}.
There, it is shown that numerical convergence is reached already
with $8^3$ particles.
}.
In the $32^3$ case also the shielding of \citet{DraineBertoldi1996}
has been adopted, with the values cited in their paper.
The initial temperature of the gas is set to $T = 10^2 \,
\rm K$ and is subject to photoheating and radiative cooling.
At a temperature of $10^4 \, \rm K$, the case-B recombination
coefficient is $\alpha_{\rm B} = 2.59 \times 10^{-13} \,\rm cm^3 \, s^{-1}$
\cite[e.g.][]{Iliev2009}.
Given these parameters, the recombination time is 
$t_{\rm  rec}=125.127 \, \rm Myr$, and the expected Str\"omgren radius 
in the isothermal case (assuming $T = 10^4 \, \rm K$) is 
$r_{\rm S} = 5.4 \,\rm kpc$.
\bfig
\centering
\includegraphics[width=0.45\textwidth]{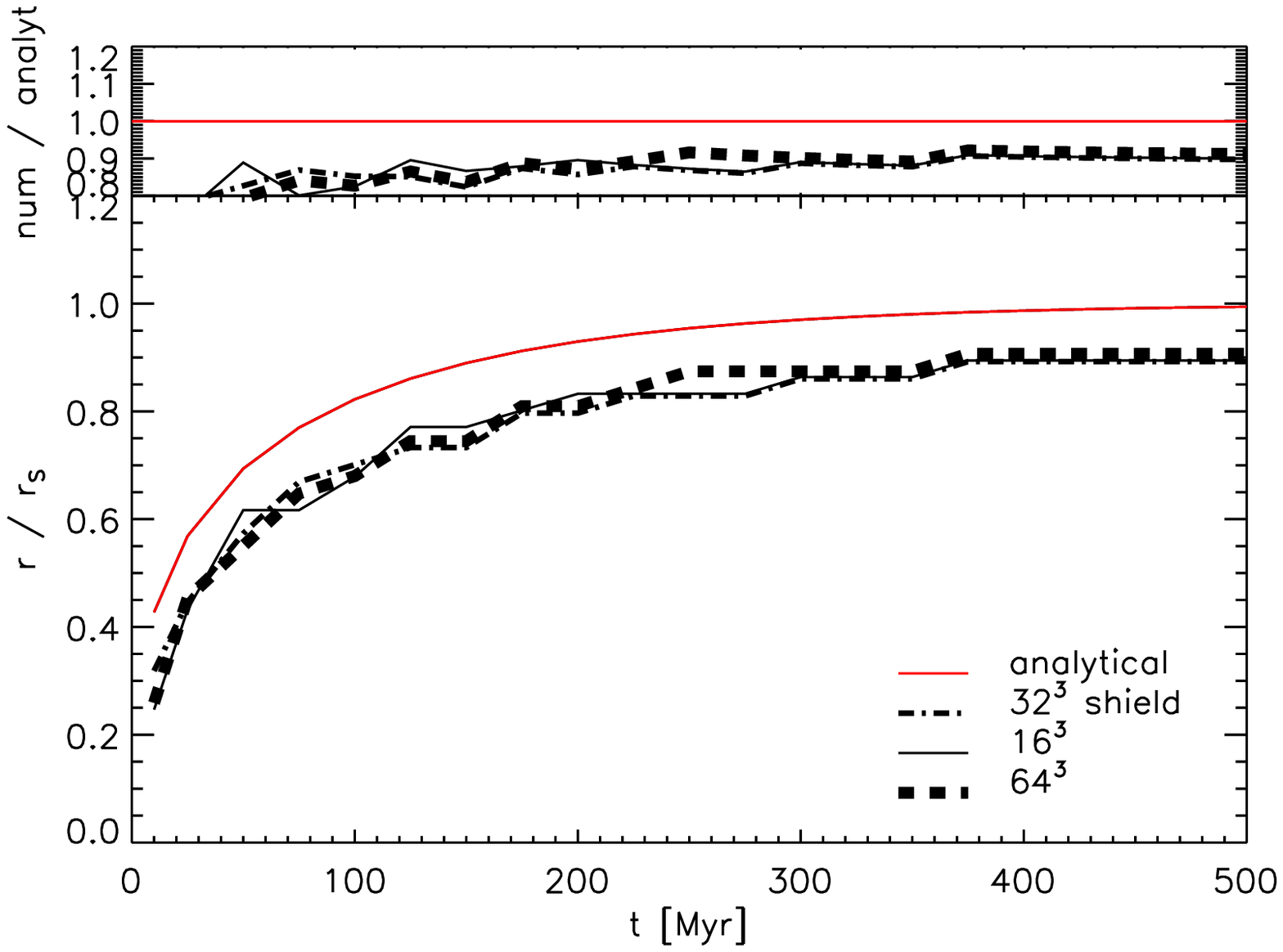}
\caption[]{\small
Evolution of the radial position of the I-front. The red line shows the analytic 
solution for an isothermal hydrogen only sphere. The black
lines show our results from the simulations, assuming the radius is
at the position where the amount of neutral and ionized hydrogen is
equal. The thick dashed line shows the results from the simulation
with $64^3$ particles, the thinner dash-dotted line -- $16^3$
particles, and the thin solid line -- $32^3$ particles, where also
H$_2$ shielding has been adopted. All lines agree very well with each
other, where the lowest resolution exhibits more scatter. The results
from the simulation with shielding show that it can be discarded in
set-ups like this, where no high particle number densities are
reached.  As shown in other studies (see text), the radius
of the I-front stays always below the analytical solution given by
eqn.~\ref{rI}.
}
\label{fig:SST_um_evol}
\efig
\bfig
\includegraphics[width=0.45\textwidth,clip]{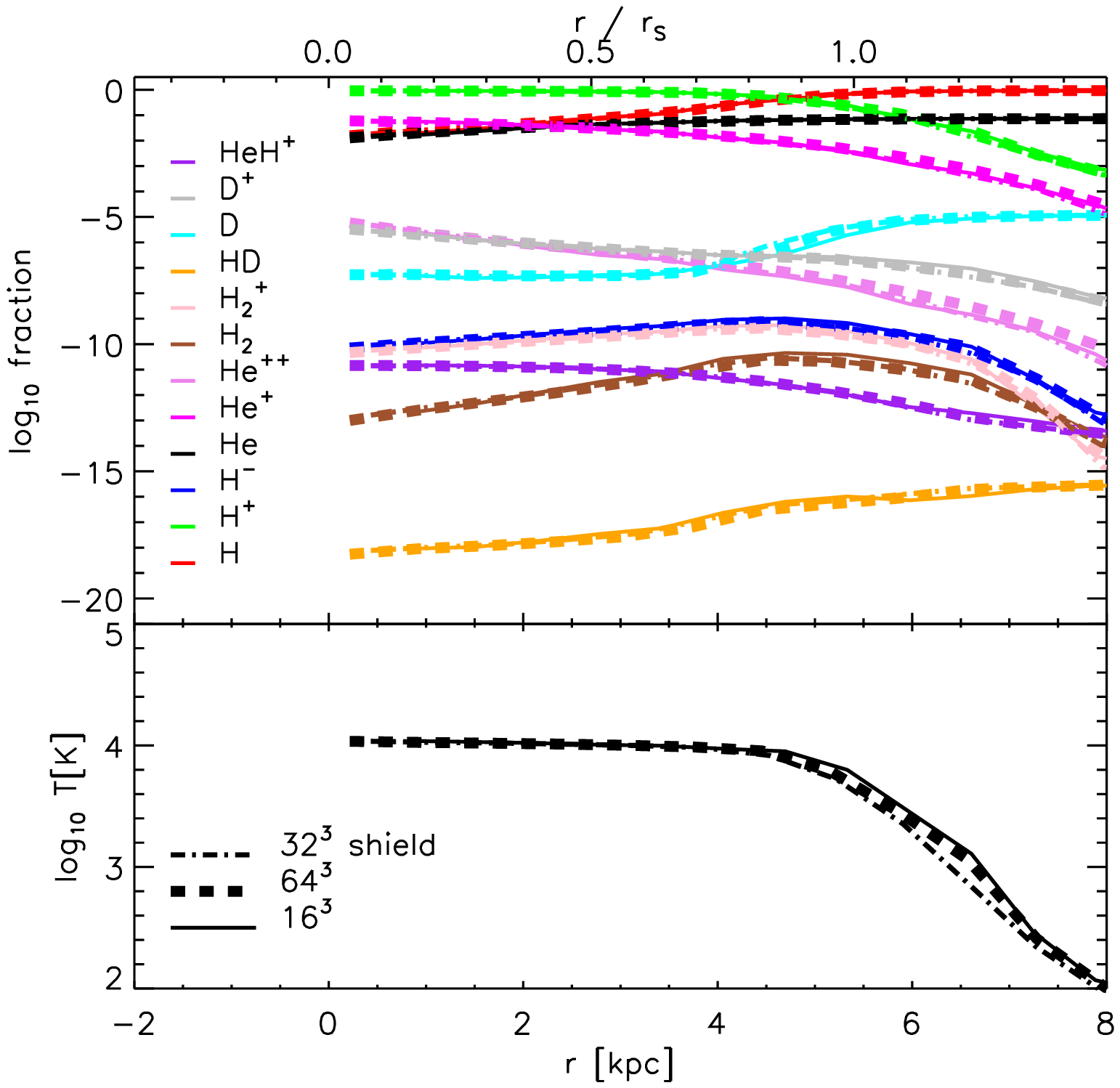}
\caption[]{\small
\textit{Top panel}: Chemical abundance fractions radial profiles at 500 Myr after
the source has been switched on. \textit{Bottom panel}: Temperature radial profiles at 500 Myr after
the source has been switched on. The temperature inside the ionized
region reaches  $\sim 10^4\, \rm K$ and extends beyond $6\, \rm
kpc$ since harder photons, unabsorbed by the gas, heat the medium ahead of the
ionization front. All simulations with different resolution ad
shielding agree very well with each other at all radii.
}
\label{fig:SST_um}
\efig
In Fig.~\ref{fig:SST_um_evol}, we show the evolution of the radial
position of the I-front with time for the different resolutions. As a proxy for the
position of the front we take the radius where the neutral and ionized
hydrogen fractions are equal (see also Fig.~\ref{fig:SST_um}).
\\
All resolutions agree very well with each other. There is no
difference in the case with shielding since the simulation never
reaches the densities required to produce some effect, as discussed
in the introduction.
Our results agree within 10\% with the analytical ones from
equation~(\ref{rI}).
In particular, the simple analytical solution is systematically larger 
than the full-simulation trend, which is expected.
This can be explained by the missing cooling contributions in
the analytical calculations from, e.g., He, H$_2$, H$_2^+$, HD that
lower temperatures, enhance recombination, and make the Str\"omgren
radius decrease (as visible in the simulated case).
In fact, equation~(\ref{rI}) is computed by assuming constant
temperature for hydrogen-only gas \cite[see also][]{Petkova2009},
while, in the numerical calculations the full chemistry treatment of
Table~\ref{tab:reactions}, including cooling and heating, is
considered. \citet{Pawlik2010} find similar results in their
one-dimensional ionized sphere simulations\footnote{Note that
  \citet{Pawlik2010} adopt a black body spectrum with effective
  temperature of $10^5$ K and therefore produce different temperature
  and He radial profiles.
}.

In Fig.~\ref{fig:SST_um}, we show the radial profile of the temperature of 
the gas at 500 Myr after the source has been switched on. The
temperature inside the ionized region reaches $\sim 10^4\, \rm K$,
consistently with photoheating from a stellar-type source, and extends
beyond $5\, \rm kpc$. 
Even further the temperature begins to drop.
Harder photons (with energies $> 60 \, \rm eV$), that do not ionize the 
elements effectively, heat the medium ahead of the ionization front.
If the source had a harder spectrum, e.g. $10^5 \, \rm K$ black body, than 
the gas would be heated even at larger radii.
We stress that at a distance of $\gtrsim 5\,\rm kpc$
(namely, around the Str\"omgren radius) temperatures steeply drop from 
$\sim 10^4\,\rm K$ down to $\sim 10^3\,\rm K$
and recombination processes take place (see next).

Correspondingly to the temperature profile, in Fig.~\ref{fig:SST_um}, 
we display the radial profiles of the different chemical abundances 
at $500\,\rm Myr$ after the radiative source has been switched on. 
We assumed initial cosmic abundances\footnote{
They are set to:
$x_{\rm e^-} \simeq 4\times 10^{-4}$,
$x_{\rm H} = 0.926$,
$x_{\rm H^+}  \simeq 4\times 10^{-4}$,
$x_{\rm H^-} = 10^{-19}$,
$x_{\rm He} = 0.07$,
$x_{\rm He^+} = 10^{-25}$,
$x_{\rm He^{++}} = 10^{-30}$,
$x_{\rm H_2} = 10^{-13}$,
$x_{\rm H_2^+} = 10^{-18}$,
$x_{\rm HD} = 10^{-16}$,
$x_{\rm D} = 10^{-5}$,
$x_{\rm D^+} = 10^{-7}$,
$x_{\rm HeH^+} = 10^{-21}$.
},
which means that hydrogen species account for $\sim 93\% $ of the total
number densities and helium species for $\sim 7\%$.
As expected, ionized fractions usually have larger
values closer to the sources (within a few kpc), while neutral or
molecular fractions increase at larger distances (above $3-6\,\rm
kpc$). In the following we discuss these trends, by referring to the
reaction network of Table~\ref{tab:reactions}, in a more precise and
detailed way.

\begin{itemize}
\item 
{\bf Atomic hydrogen species:} due to the strong radiation intensities
near the central source, hydrogen is kept completely ionized
(i.e. the total hydrogen fraction, $\sim 0.93$, is in the ionized
state, H$^+$) within $r\lesssim 5 \,\rm kpc$ by the dominant
photoionization process 
\be
\label{Hion}
\rm H + \gamma \rightarrow H^+ + e^{-}.
\ee
Only at larger radii, where the radiation intensity decreases, the recombination process
\be
\label{Hrec}
\rm H^+ + e^{-} \rightarrow H + \gamma
\ee
takes over and makes the H$^+$ fraction drop down of several orders of
magnitude, with the consequent increase of H fraction up to $\sim
0.93$ (namely, the total hydrogen fraction is now in the neutral
state). 
\\
H$^-$ is a very important species because it represents one of the
main channels by which molecular hydrogen can be formed (see
discussion later). 
It shows fractional values of $\sim 10^{-9}-10^{-10}$ until $r\lesssim
6\,\rm kpc$ and at larger distances drops of roughly $3-4$ orders of
magnitude. 
The increment of H$^-$ fraction up to $\sim 10^{-9}$ in
correspondence of $r\sim 4-5\,\rm kpc$ is due to the simultaneous
hydrogen recombination -- eq. (\ref{Hrec}) -- which leads to an
increase of the neutral hydrogen and enhances the H$^-$ formation via 
\be
\label{Hminus_creation}
\rm H + e^{-} \rightarrow H^- + \gamma
\ee
reaction.
In the innermost regions, due to the low ionization energy of
only $0.755~$eV, H$^-$ photoionization 
\be
\label{Hminus_photoion}
\rm H^- + \gamma \rightarrow H + e^-
\ee
and collisional destruction by the abundant H$^+$ and $e^-$ species
\be
\label{Hminus_destruction_Hplus1}
\rm H^- + H^+ \rightarrow 2H
\ee
\be
\label{Hminus_destruction_Hplus2}
\rm H^- + H^+ \rightarrow H_2 + e^-
\ee
\be
\label{Hminus_destruction_e}
\rm H^- + e^- \rightarrow H + 2e^-
\ee
determine a lower fraction of $\sim 10^{-10}$.
At larger distances ($r\sim 5-10\,\rm kpc$), H is dominant, but, contrary to reaction (\ref{Hminus_creation}), free $e^-$ are lacking and the most effective reactions are
\be
\label{Hminus_destruction_H1}
\rm H^- + H \rightarrow H_2 + e^-
\ee
\be
\label{Hminus_destruction_H2}
\rm H^- + H \rightarrow 2H + e^-
\ee
that lead to a decrease of H$^-$ fraction down to $\lesssim 10^{-13}$.

We note that a calculation of the exact analytical expression for the
the Str\"omgren radius and a comparison with our results is not
meaningful, since the analytical study is based on the simplified case
of hydrogen-only gas and does not take into account interaction with
other species and the effects of
photoheating. However, the radius of the ionized hydrogen reaches
$\sim 5 \, \rm kpc$, which is approximate to the expectation value
of the Str\"omgren radius for the isothermal case. 

\item
{\bf Atomic helium species:} in the same way, due to the reactions
\be
\label{He_destruction1}
\rm He + \gamma \rightarrow He^+ + e^-
\ee
\be
\label{He_destruction2}
\rm He + e^- \rightarrow He^+ + 2e^-
\ee
neutral He is efficiently destroyed at $r\lesssim 3\,\rm kpc$, and
the residual fraction is $\sim 0.01$.
He$^+$ and He$^{++}$ reach fractions of $\sim 0.06$ and 
$\sim 10^{-5}$, respectively, via 
\be
\label{Heplus_destruction1}
\rm He^+ + e^- \rightarrow He^{++} + 2e^-
\ee
\be
\label{Heplus_destruction2}
\rm He^+ + \gamma \rightarrow He^{++} + e^-
\ee
\be
\label{Heplusplus_destruction1}
\rm He^{++} + e^- \rightarrow He^+ + \gamma.
\ee
Moreover, the additional He depletion by H$^+$ collisions leads to the
formation of HeH$^+$ (as we will discuss later). 
At $r > 3\,\rm kpc$, the decreasing intensity of ionizing radiation --
which plays the most relevant role in
equations~(\ref{He_destruction1}) and (\ref{Heplus_destruction2}) --
and the 
ongoing recombination processes which take away free electrons from
the medium -- needed e.g. in reactions (\ref{He_destruction2}),
(\ref{Heplus_destruction1}) --  cannot sustain ionization any longer
and the trends exhibit monotonic radial drops for both He$^+$ and
He$^{++}$. 
More exactly, He$^+$ and He recombine according to  
\be
\label{Heplus_rec1}
\rm He^{++} + e^- \rightarrow He^+ + \gamma
\ee
\be
\label{He_rec1}
\rm He^{+} + e^- \rightarrow He + \gamma
\ee
respectively, so, the final states of He$^{+}$ and He$^{++}$ are
strictly linked to each other with abundances of $\sim 10^{-5}$ and
10$^{-11}$ at $r\sim 8\,\rm kpc$. 
Radiative interactions are weaker and weaker and they become
practically negligible at large radii. 

\item
{\bf Atomic deuterium species:} the behaviour of D and D$^+$ are
quite regular and similar to H and H$^+$, with D$^+$ being dominant at
$r \lesssim 5\,\rm kpc$ and D dominant at $r> 5\,\rm kpc$ of $\sim 3$
orders of magnitude. 
The abundances of deuterium and hydrogen species are bound by the 
balance reactions 
\be
\label{D1}
\rm H^{+} + D \rightarrow H + D^+
\ee
\be
\label{D2}
\rm H + D^+ \rightarrow H^+ + D ,
\ee
but the most efficient processes are still photoionization (`close' to the source) 
\be
\label{D_photoion}
\rm D + \gamma \rightarrow D^+ + e^-
\ee
and recombination (`far' from the source)
\be
\label{D_rec}
\rm D^+ + e^- \rightarrow D + \gamma.
\ee
Minor contributions of D and D$^+$ are also involved for molecular
species (see next) and can slightly affect H or H$^+$ production. 

\item
{\bf Molecular species:} Finally, we discuss the trends of molecular
species (H$_2$, H$_2^+$, HD, HeH$^+$). 
Their profiles are less regular and intuitive than atomic profiles,
since more processes need to be addressed at the same time. 
For example, H$_2$ and H$_2^+$ have fractional values of $\sim
10^{-13}$ and $\sim 10^{-10}$, respectively, at $r\sim 1\,\rm kpc$,
then, they show an increasing trend and a peak of $\sim
10^{-10}-10^{-9}$ at $r \sim 4-6\,\rm kpc$. 
For larger $r$, they exhibit a drop off below $\sim 10^{-14}$, with H$_2$ overcoming H$_2^+$ at $r>7\,\rm kpc$.
These behaviours are understood by considering that in the innermost
regions the radiative negative feedback on molecule formation
decreases for increasing $r$: this means that at larger radii
radiation is not strong enough to dissociate molecules via 
\be
\label{H2_destruction}
\rm H_2+ \gamma \rightarrow 2 H
\ee
and, thanks to the available H, H$^+$ and $e^-$ (simultaneously
present at $r\sim 4-6\,\rm kpc$, where H recombination is still taking
place), H$_2$ formation can proceed through the H$^-$ channel -- see
also reaction~(\ref{Hminus_creation}), 
\be
\rm H + e^- \rightarrow H^- + \gamma, \nonumber
\ee
\be
\rm H^- + H \rightarrow H_2+e^-,
\ee
and through the  H$_2^+$ channel,
\be
\rm H + H^+ \rightarrow H_2^+ + \gamma\\
\ee
\be
\rm H_2^+ + H \rightarrow H_2+H^+.
\ee
This is the intrinsic reason why molecular hydrogen roughly follows
the H$^{-}$ profile discussed earlier. 
Obviously, residual photons will slightly boost the ionized fraction
of H$_2^+$ (with respect to H$_2$) via 
\be
\rm H_2 + \gamma \rightarrow H_2^+ + e-
\ee
until $r\sim 7\,\rm kpc$.
At larger distances, photons are too weak to ionize the gas, so H$_2$
takes over and H$_2^+$ drops dramatically of $\sim 3$ orders of
magnitude within $1\,\rm kpc$. 
Additionally, we note that the sharp decrement of molecular fractions
at very large distances is basically due to the fact that the medium
becomes almost completely neutral and the ionized fractions of $e^-$
and H$^+$ are too low to boost H$^-$ and H$_2^+$, and, hence, H$_2$
formation. 

Similarly, the increase with radius of HD fraction is essentially
caused by the weakening of the central radiation and of H$^+$ and
e$^-$ fractions which are less and less effective  in dissociating it
at larger $r$.
In particular, at $r\lesssim 4\,\rm kpc$, H, D and H$_2$ are
subdominant with respect to their ionized counterparts H$^+$, D$^+$
and H$_2^+$, thus, HD formation due to
\be
\label{HDformation1}
\rm D + H_2 \rightarrow HD + H
\ee
is strongly inhibited, while little contributions come from
\be
\label{HDformation2}
\rm D^+ + H_2 \rightarrow HD + H^+.
\ee
Destruction from reaction
\be
\label{HDdestruction}
\rm HD + H^+ \rightarrow D^+ + H_2
\ee
further lower HD abundances around $\sim 10^{-18}$ level.
At $r\gtrsim 4\,\rm kpc$, hydrogen and deuterium recombinations, 
together with H$_2$ formation via H$^-$ channel, support HD formation 
through reactions~(\ref{HDformation1}) and (\ref{HDformation2}), instead
reaction~(\ref{HDdestruction}) is no longer effective.
As a consequence, the HD fractional abundance grows more than $\sim 2$ 
orders of magnitude at $r\sim 8\,\rm kpc$.
\\
We note that a boost of HD production in shock-compressed gas was
expected because of the simultaneous increment of H$_2$ fractions, 
quite visible in Fig.~\ref{fig:SST_um}, in correspondence of the 
Str{\"o}mgreen radius.

For what concerns the aforementioned HeH$^+$, this is efficiently
produced near the source because there are a lot of free protons
which can boost its abundance via 
\be
\rm He+H^+ \rightarrow HeH^+ + \gamma
\ee
even in a more powerful way than photodissociation
\be
\rm HeH^+ + \gamma \rightarrow He + H^+.
\ee
Only when protons are lacking (i.e. at large $r$) HeH$^+$ production
is inhibited up to $\sim 3$ orders of magnitude and drops from $\sim
10^{-11}$ to $\sim 10^{-14}$. 
\end{itemize}

We stress that the residual relative ionized fractions far from the
source ($r\sim 8\,\rm kpc$) are: 
$n_{\rm H^+}/n_{\rm H}\sim 10^{-3}$,
$n_{\rm D^+}/n_{\rm D}\sim 10^{-3}$,
$n_{\rm H^-}/n_{\rm H}\sim 10^{-13}$,
$n_{\rm He^+}/n_{\rm He}\sim 10^{-3}$,
$n_{\rm He^{++}}/n_{\rm He}\sim 10^{-9}$,
$n_{\rm H_2^+}/n_{\rm H_2}\sim 10^{-1}$.
The absolute values quoted in the previous discussion are dependent on
the initial composition assumed for the gas. 
Although the qualitative behaviour is not supposed to change much,
larger or smaller values for the assumed fractions could result in
more efficient or less efficient formation and destruction processes. 
However, the general trends  are supposed to be quite independent from that.
\\

The test presented here is based on the initial conditions of 
\cite{Iliev2009}, but different works are available in literature.
Results that are going in the same direction were in fact found by
e.g. \cite{Ricotti2001, Ahn2007, WhalenNorman2008}, who developed,
independently, different H, He, and H$_2$ chemistry networks.
Their simulation setups were very different, though, and the parameters
for the central source, as well.
\\
\cite{Ricotti2001} (their Fig.~3) performed the first one-dimensional
calculations of a star shining in a primordial mini-halo.
A popIII-like source with a power-law spectrum and 
$\dot N_\gamma\simeq 1.2\times 10^{49}\,\rm s^{-1}$ at $z\simeq 19$
was located in a static, uniform medium at the mean density
$ \bar{n}(z\simeq19) \simeq 0.1\,\rm cm^{-3}$.
They studied the behaviour of the molecular species after $\sim 100\,\rm Myr$ 
from the explosion of the central star, when the Str{\"o}mgreen radius
had reached $\sim 4\,\rm kpc$, and a corresponding peak of H$_2$ with
$x_{\rm H_2}\sim 5\times 10^{-4}$, a few kpc wide, had formed.
This was equivalent to a boost of about $3$ orders of magnitude
with respect to their initial $x_{\rm H_2}\sim 10^{-6}$ value assumed.
Also in our previous discussion, despite the different initial values,
we have found an increase in the H$_2$ fraction of $\sim 3$ orders of
magnitude.
\\
Similarly,  \cite{Ahn2007} (their Fig.~8) and \cite{WhalenNorman2008} 
(their Fig.~1) performed one- and three-dimensional calculations, respectively, 
considering a $120\,\rm \msun$ popIII-like source with a blackbody spectrum having 
effective temperature of $T_{eff}=10^5\,\rm K$ and 
$\dot N_\gamma \simeq 1.5\times 10^{50}\,\rm s^{-1}$.
They assumed the star to be in a primordial halo with a truncated isothermal 
sphere density profile, whose central matter densities were reaching up 
$\gtrsim 4\times 10^{-22}\,\rm g\,cm^{-3}$.
% (i.e. $\sim 10^{4}$ the mean cosmic gas density).
They focused on the innermost core of the star forming regions, after roughly
0.6 times the lifetime of the massive popIII stars (a few Myr), when the 
Str{\"o}mgreen radius was still around $\lesssim 30\,\rm pc$ and a corresponding 
peak of H$_2$ $\sim 10~$pc wide had arisen.
They also find an increase to $x_{\rm H_2}\sim 10^{-4}$, a few orders of magnitude
larger than the assumed initial value of $x_{\rm H_2}\sim 10^{-6}$.
\\
The first relevant difference with the study presented here is in the assumed
$\dot N_\gamma$ (our value is $5\times 10^{48}\,\rm s^{-1}$):
\cite{Ricotti2001} used an $\dot N_\gamma $ 2.2 times larger, 
while \cite{Ahn2007} and \cite{WhalenNorman2008} used a value 300 times larger
Consequently, the number of photons available to ionize atoms or dissociate
molecules is much bigger in their cases, than in ours.
This explains why they find H$_2$ fractions more significantly destroyed in the
inner regions.
\\
Then, molecule dissociation or creation is also strictly linked to the 
typical conditions of the medium.
In the aforementioned works, this is denser than in ours (of $\sim 2 - 4$ orders 
of magnitude), and have strong implications when computing abundances, since higher
densities allow easier collisional dissociation in the inner, ionized regions.
When temperatures fall below $\sim 10^4\,\rm K$, H$_2$ creation becomes more 
efficient and larger free-electron number densities can lead to a noticeable increase 
of molecular fractions\footnote{
  For example, electron fractions of $\sim 10^{-4}$ determine typical IGM fractions
  for H$_2$ of $\sim 10^{-6}$, instead, in gas with overdensities of $\sim 10^4$,
  H$_2$ fractions can be boosted up to $\sim 10^{-3}$ \cite[][]{Ahn2007}.
}.
This is actually the reason why the profiles of our low-density gas in
Fig.~\ref{fig:SST_um} show H$_2$ increases up to only $\sim 10^{-10}$,
while very overdense gas can reach H$_2$ fractions of $\sim 10^{-4}$ in 
front of the propagating I-front.
\\
We conclude by noticing that, despite the huge diversities in the parameters and in
the H$_2$ peak values, the different behaviours of the profiles are in qualitative 
agreement, showing an increment of the molecular fractions in correspondence of the 
I-front, and orders-of-magnitude declines farther away.
Together with the H$_2$ raise (within a shell of $\sim 2\,\rm kpc$) the consequent HD 
increment is obviously expected to be more prominent in high-density environments.
\\
None the less, the final effects on triggering star formation are not completely
clear, since gas runaway collapse can be significantly enhanced only when densities 
are larger than $\sim 1-10\,\rm cm^{-3}$, and molecular fractions increase up 
to $> 10^{-2}$ (see further discussions in Sect.~\ref{Sect:cosmo}).

\bfig
\centering
\includegraphics[width=0.45\textwidth]{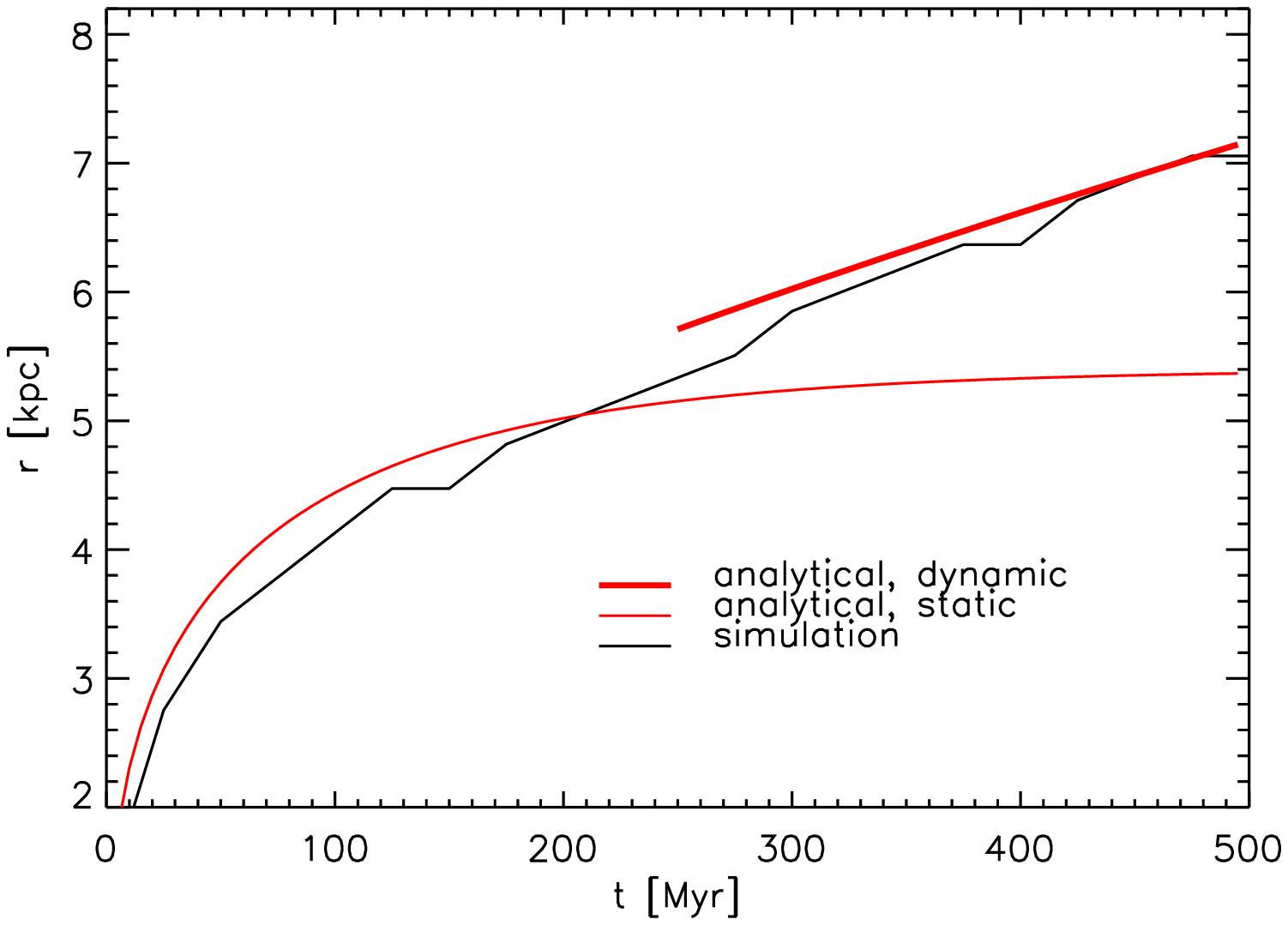}
\caption[]{\small
Evolution of the radial position of the I-front. The dashed
line shows our results from the simulation, assuming the radius is
at the position where the amount of neutral and ionized hydrogen is
equal. The solid line is the result from eq.~(\ref{rI}), 
computed by assuming isothermal gas at $10^4\,\rm K$. The extension of
the solid line is given by eq.~(\ref{rI,d}). The results from the
simulation agree very well with the analytical results, also at radii
beyond the Str\"omgren radius $r_{\rm S}$.
}
\label{fig:SST_um_dyn_evol}
\efig

\bfig
\includegraphics[width=0.45\textwidth,clip]{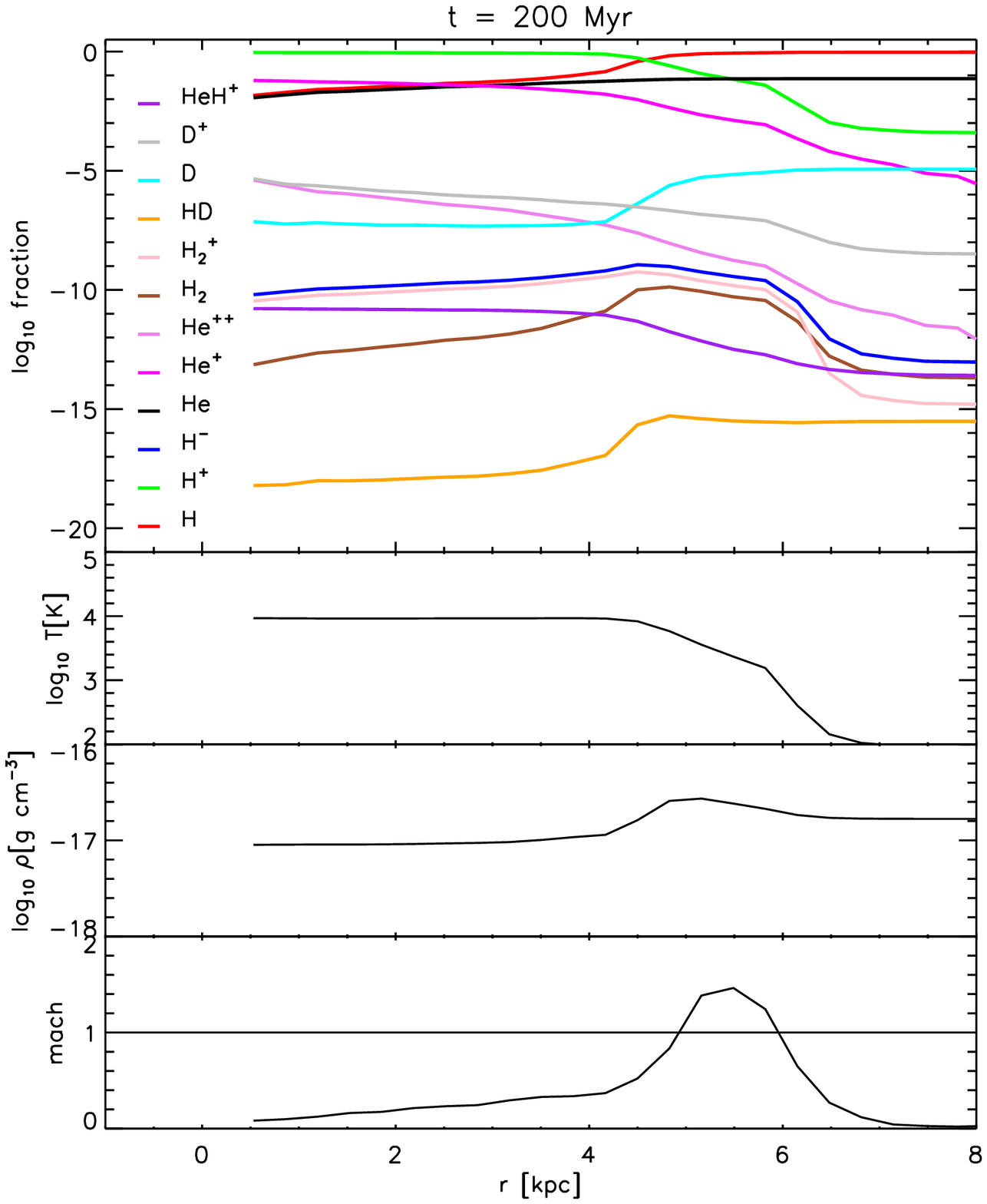}
\caption[]{\small
\textit{Top panel}: Chemical abundance fractions radial profiles at 500 Myr after
the source has been switched on. \textit{Bottom panel}: Temperature,
density, and Mach number radial profiles at $200$ Myr after
the source has been switched on. The temperature inside the ionized
region reaches  $\sim 10^4\, \rm K$ and extends beyond $6\, \rm
kpc$ since harder photons, unabsorbed by the gas, heat the medium ahead of the
ionization front. A shock, a H$_2$ shell, and an HD boost have
developed ahead of the I-front.}
\label{fig:SST_um_dyn}
\efig

\subsubsection{Ionized sphere expansion in a dynamic density field}\label{sec:dyn}
In this section we repeat the test from the previous one, but we allow
the gas particles to move due to pressure forces, but not gravity. The density field is not static
anymore. The resolution we have chosen here is $32^3$ gas particles. 

The position of the I-front in 
this stage is given by \citep{Spitzer1978} \be r_{\rm I} = r_{\rm                                                         
  S}\left( 1+\frac{7c_{\rm s}t}{4r_{\rm                                                                                 
    S}}\right), \label{rI,d}\ee where $c_{\rm s}$ is the sound speed
of the ionized gas and $r_{\rm S}$ is the Str\"omgren radius given
by equation~(\ref{rI}).

In Fig.\ref{fig:SST_um_dyn_evol} we show the evolution of the position of the
I-front and its velocity with time. We compare with analytical results
from equations ~(\ref{rI}) and ~(\ref{rI,d}). We note that the I-front follows the analytical
prediction for static gas in the beginning of the expansion. After
approximately $200 \, \rm Myr$ at approximately $5\, \rm kpc$ from
the source, the I-front continues to move outwards, rather than start
to decelerate, and follows the dynamic gas solution given above.

In Fig.~\ref{fig:SST_um_dyn} we present the radial profiles of the different
species, the temperature, and the density of the gas at $\rm t = 200
\, Myr$ after the source has been switched on. The temperature in the
ionized sphere is $\sim 10^4 \rm \, K$ and there is a shock and a density
contraction ahead of the I-front, which moves at super-sonic speed at
this evolution time. We also note that there is a HD boost and a H$_2$ shell ahead of the
I-front as well, inside the density contraction region. Similar
results have been shown also by \citet{Ricotti2001}. These
conditions -- increased density and increased H$_2$ fraction are
favorable for star formation.

%************************************************************************

\subsection{Cosmological abundance evolution}\label{Sect:abundances}
\bfig
\centering
\includegraphics[width=0.45\textwidth,clip]{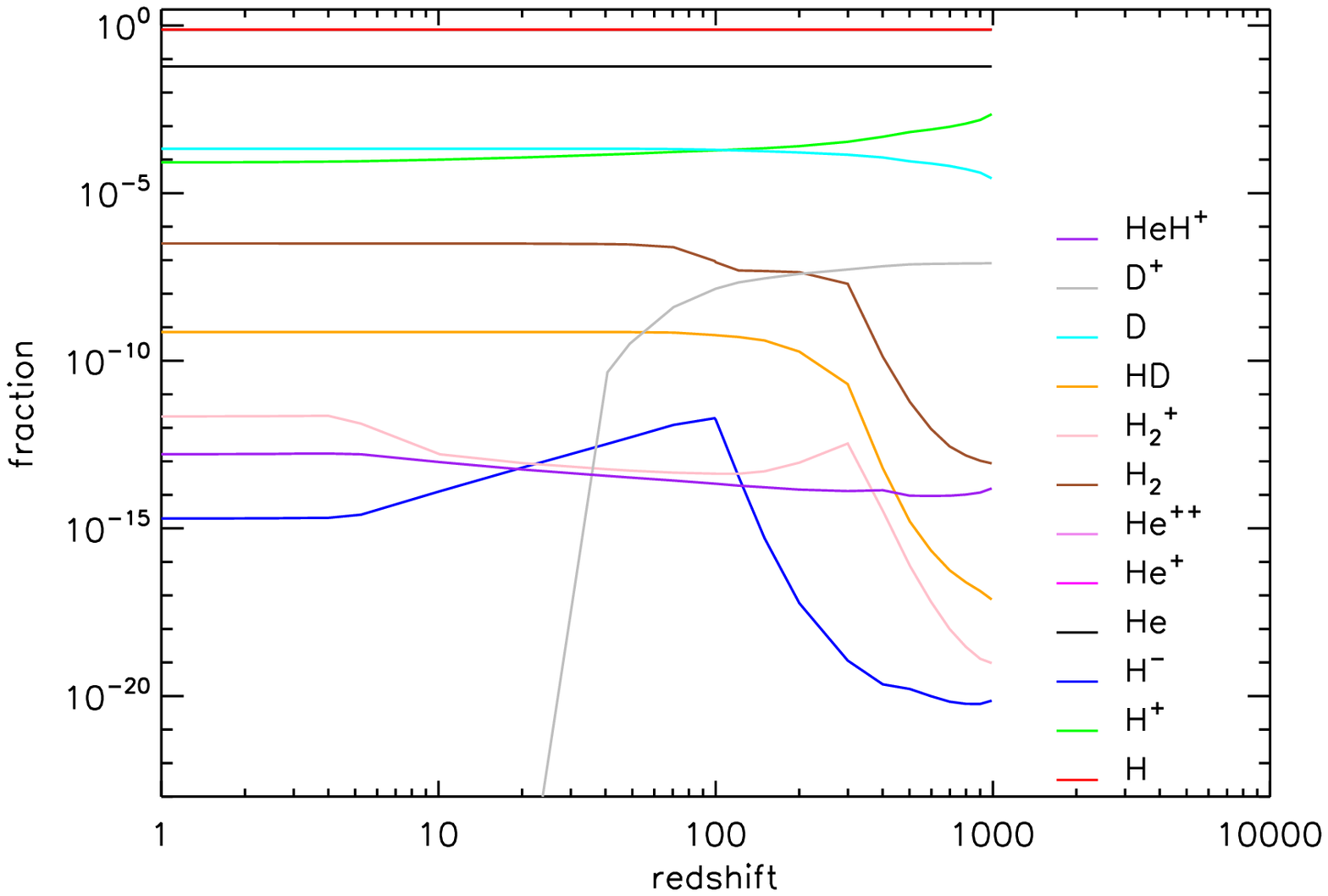}
\caption[]{\small
Number fraction evolution as a function of redshift for the
different chemical species, followed in our implementation. The
background cosmology is a standard flat $\Lambda$CDM model with
geometrical parameters:
$\Omegat = 1.0$, $\Omegal = 0.7$, $\Omegam = 0.3$, $\Omegab = 0.04$.} 
\label{fig:cosmo_evolution}
\efig
In order to test how our implementation performs during
cosmological evolution, we run the chemical network coupled with the
RT network for the mean background density evolution
and follow the changes in the different species as a function of the
cosmic time. 
The radiative source is assumed to be the uniform CMB radiation with a
black body spectrum with effective temperature of $\sim 2.73 \,
(1+z)\,\rm K$. The CMB radiation is 
self-consistently followed, according to the treatment outlined in
Sect.~\ref{Sect:implementations}.
The initial fractions, $x$, for the different species\footnote{
They are set to:
$x_{\rm e^-} \simeq 4\times 10^{-4}$,
$x_{\rm H} = 0.926$,
$x_{\rm H^+}  \simeq 4\times 10^{-4}$,
$x_{\rm H^-} = 10^{-19}$,
$x_{\rm He} = 0.07$,
$x_{\rm He^+} = 10^{-25}$,
$x_{\rm He^{++}} = 10^{-30}$,
$x_{\rm H_2} = 10^{-13}$,
$x_{\rm H_2^+} = 10^{-18}$,
$x_{\rm HD} = 10^{-16}$,
$x_{\rm D} = 10^{-5}$,
$x_{\rm D^+} = 10^{-7}$,
$x_{\rm HeH^+} = 10^{-21}$.
} are initialized, at redshift $z\simeq 10^3$, accordingly to a
neutral plasma at $\sim 10^3\,\rm K$ \cite[see e.g.][]{GP98}. 

We present the results in Fig.~\ref{fig:cosmo_evolution}, where the
cosmic mean number fractions as a function of redshift are plotted for
a standard flat $\Lambda$CDM cosmology with geometrical parameters
$\Omegat = 1.0$,
$\Omegal = 0.7$,
$\Omegam = 0.3$,
$\Omegab = 0.04$.
The H and He number fractions are unaffected by the redshift evolution.

At early times, some residual recombination process continue taking
place, while the CMB temperature goes down, and make H$^+$ evolution
drop from a fraction of $\sim 10^{-3}$, at $z\sim 1000$, to the final
$\lesssim 10^{-4}$ value.
The evolution of H$^-$ clearly shows the effects of free electrons at
high redshift that boost its formation (and H-derived molecule
formation) from $x_{\rm H^-}\sim 10^{-20}$, at $z\sim 1000$, to
$x_{\rm H^-}\sim 10^{-12}$, at $z\sim 100$. 
The following hydrogen recombination implies a decrease both in H$^+$
and $e^-$ fraction, with consequent drop of H$^-$ (see
Table~\ref{tab:reactions}) down to $x_{\rm H^-}\sim 10^{-15}$ at low
redshift.
The two ionization states of He are constantly kept at very low
values, close to the initial ones (their trends are not displayed in
Fig.~\ref{fig:cosmo_evolution} for sake of clarity). 

As already mentioned, hydrogen molecule formation is initially
enhanced due to the available residual $e^-$ and the progressively
diminishing effects of CMB radiation: this allows hydrogen to
increasingly form H$_2^+$, with a peak around $z\sim 300$, and H$_2$,
until $z\sim 70$. 
H$_2^+$ formation is efficiently driven by H and the available H$^+$ in primordial times.
At $z \lesssim 300$, paucity of free protons (whose fraction in the
meantime has dropped of about one order of magnitude) make difficult
H$_2^+$ formation. 
However, the continuous increment of H$^-$ enhances H$_2$ at $z\sim
100$, and afterward its formation is mainly driven by the H$^-$
channel, rather than the  H$_2^+$ one, until $z\sim 70$. 
The dominant formation path at different times is clearly recognizable
when comparing the H$_2$ trend with the H$_2^+$ and H$^-$ trends. 
The $z\sim 300$ peak of H$_2^+$ corresponds to the steep increase of
$x_{\rm H_2}$ at early times, while the $z\sim 100$ peak of H$^-$
corresponds to the following boost at later times. 
For $z\lesssim 70$, $x_{\rm H_2}$ stays roughly constant between
$10^{-7}-10^{-6}$ because further production is halted by the
decrement of free protons and electrons, and radiative destruction
cannot take place because of the low CMB flux at low redshift.

Deuterium is also affected by cosmological evolution and
the weaker CMB intensity effects at lower redshift. 
Thus, recombination of D and D$^+$ make $x_{\rm D}$ increase up to $\sim
10^{-4}$ and $x_{\rm D^+}$ dramatically drop down at $z<100$. 
The simultaneous ongoing H$_2$ formation at $z \gtrsim 70$ also
`drags' HD fractions up to $\sim 10^{-9}$ levels (HD is very
sensitive to H$_2$ abundances). 

The HeH$^+$ molecule is often formed behind fast shocks
\cite[e.g.][]{NeufeldDalgarno1989} by He and H$^+$, but, due to its
low dissociation energy
\cite[of $\sim 14873.6 \,\rm cm^{-1}\simeq 1.7\,\rm eV$;][]{BishopCheung1979} 
it can be found and emitted only below $\sim 10^4\,\rm K$.
Moreover, the presence of background radiation can dissociate it in
its two components soon after the recombination epoch, at $z\lesssim
10^3$. 
Indeed, the plot in Fig.~\ref{fig:cosmo_evolution} highlights how
HeH$^+$ is initially underabundant and then is gradually formed while
the Universe expands, cools and the CMB radiation gets weaker. 
Over the cosmic time $x_{\rm HeH^+}$ slightly increases of one order
of magnitude, from $\sim 10^{-14}$ up to $\gtrsim 10^{-13}$, and, due
to H interactions, it sustains H$_2^+$ with subsequent H$_2$
formation.

The low radiation intensity of the CMB in not able to produce large
changes in the abundances of the elements, but is never the less a
good test on the performance of our implementation.
Our results agree very well with previous cosmological abundance
evolution studies as e.g. \cite{Abel_et_al1997,GP98,Maio2007}.

The presence of an additional cosmic UV radiation during reionization
\cite[e.g.][]{HaardtMadau1996} at low redshift ($z\lesssim 7$), would
heat the medium and change the ionization equilibria. 
As a consequence, one might expect more free electrons, larger H$_2^+$
and H$^-$ abundances and hence more H$_2$ production, accompanied by
increased D$^+$ and HD fractions, and dissociation of HeH$^+$.

%************************************************************************

\section{Application: cosmological structure formation}\label{Sect:cosmo}
\bfigs
\centering
\includegraphics[width=0.35\textwidth]{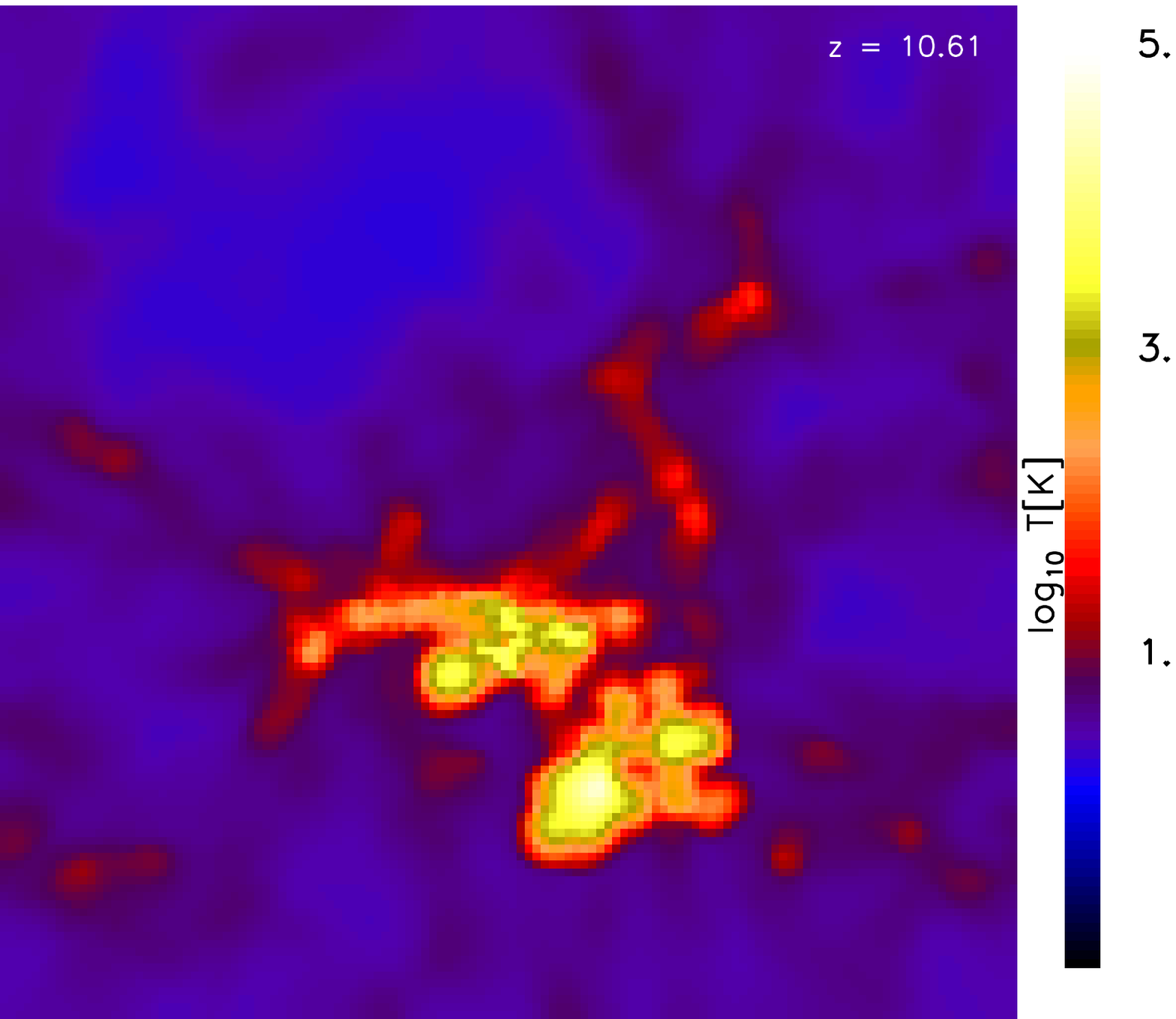}
\hspace{1.0cm}
\includegraphics[width=0.35\textwidth]{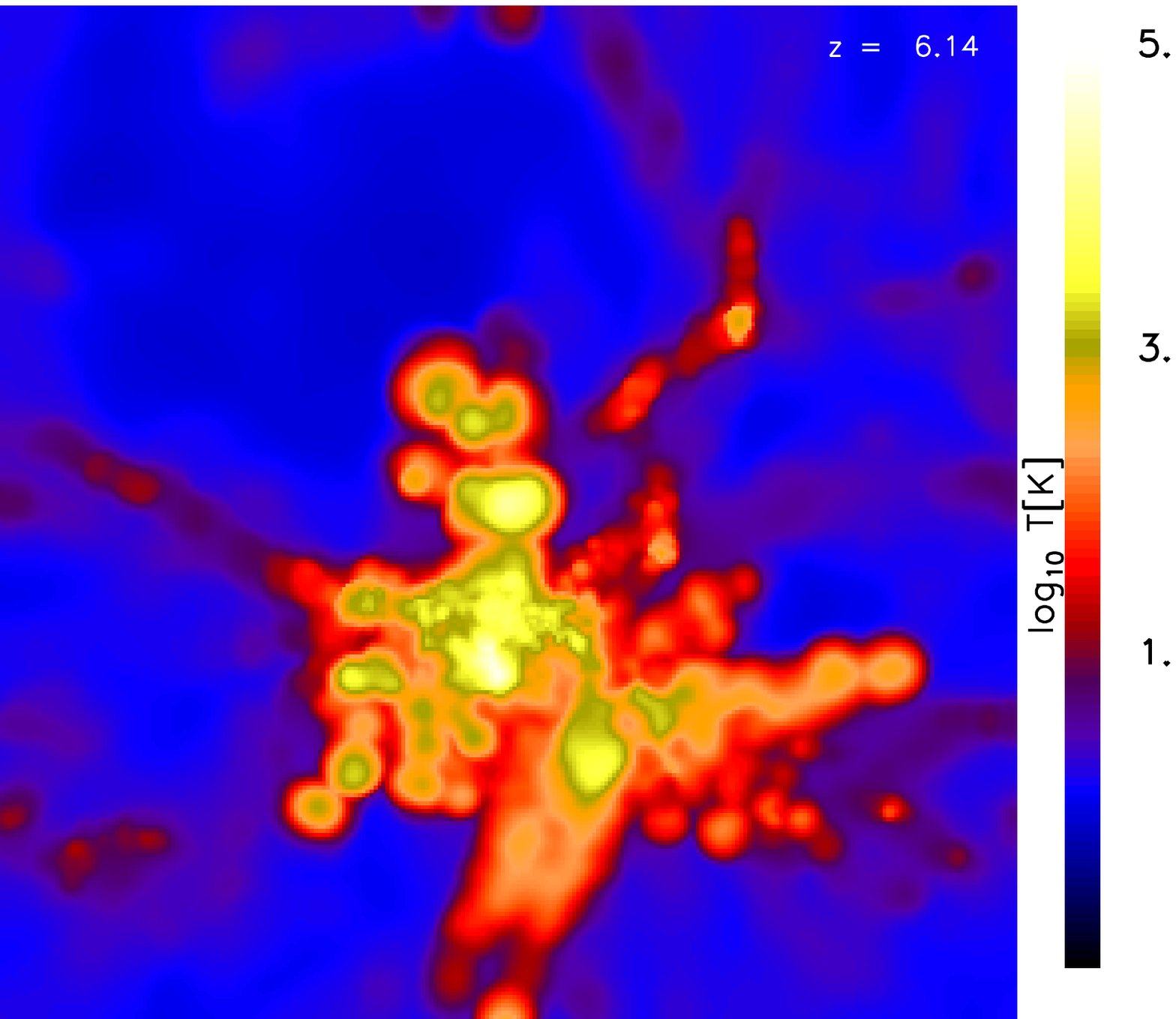}
\caption[]{\small
Mass-weighted temperature slices through the box at
redshift $z=10.61$ (left) and $z=6.14$ (right).
The simulation includes feedback effects,
full non-equilibrium chemistry and RT (see text).
}
\label{fig:maps}
\efigs
\bfigs
\centering
\includegraphics[width=0.4\textwidth]{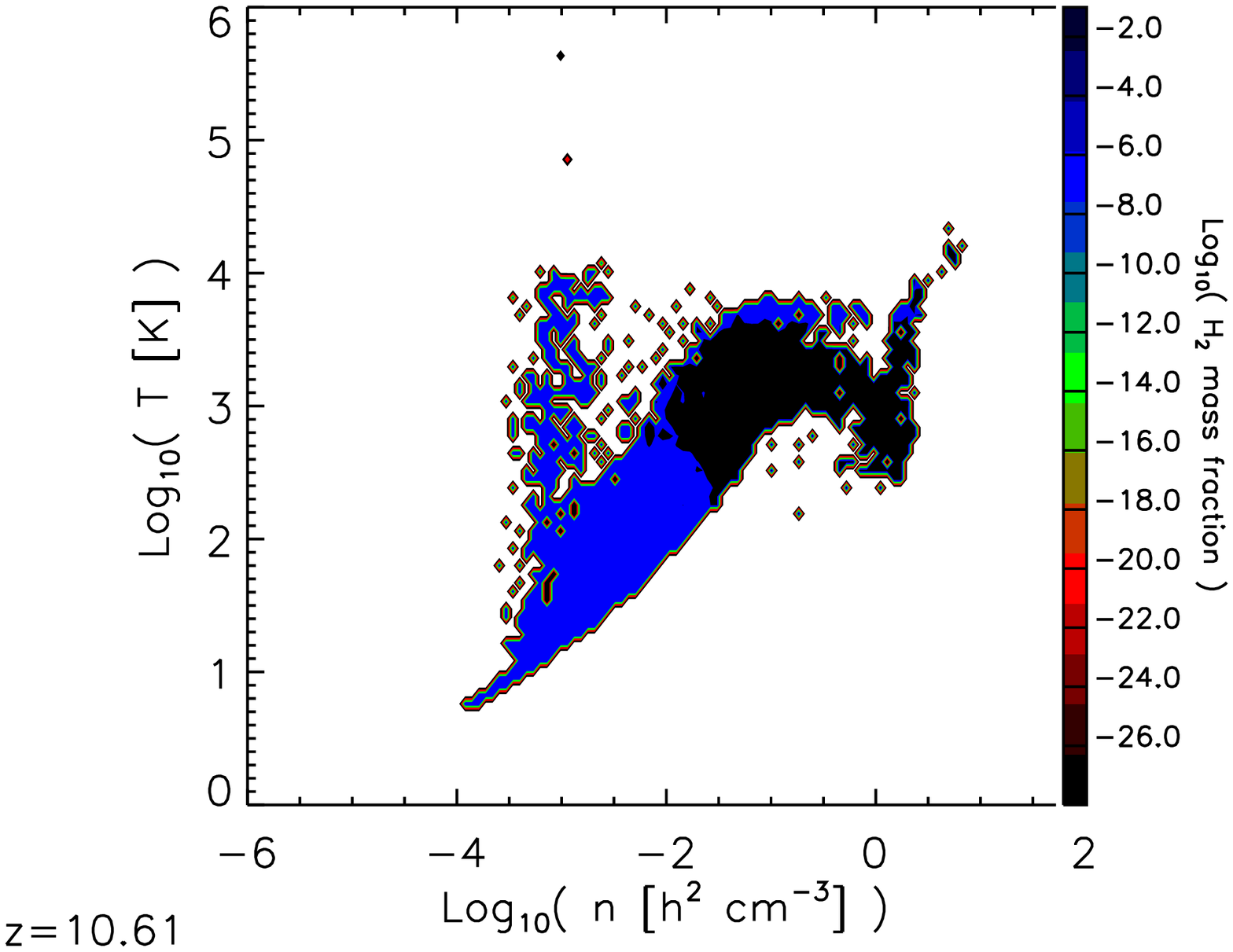}
\hspace{1.0cm}
\includegraphics[width=0.4\textwidth]{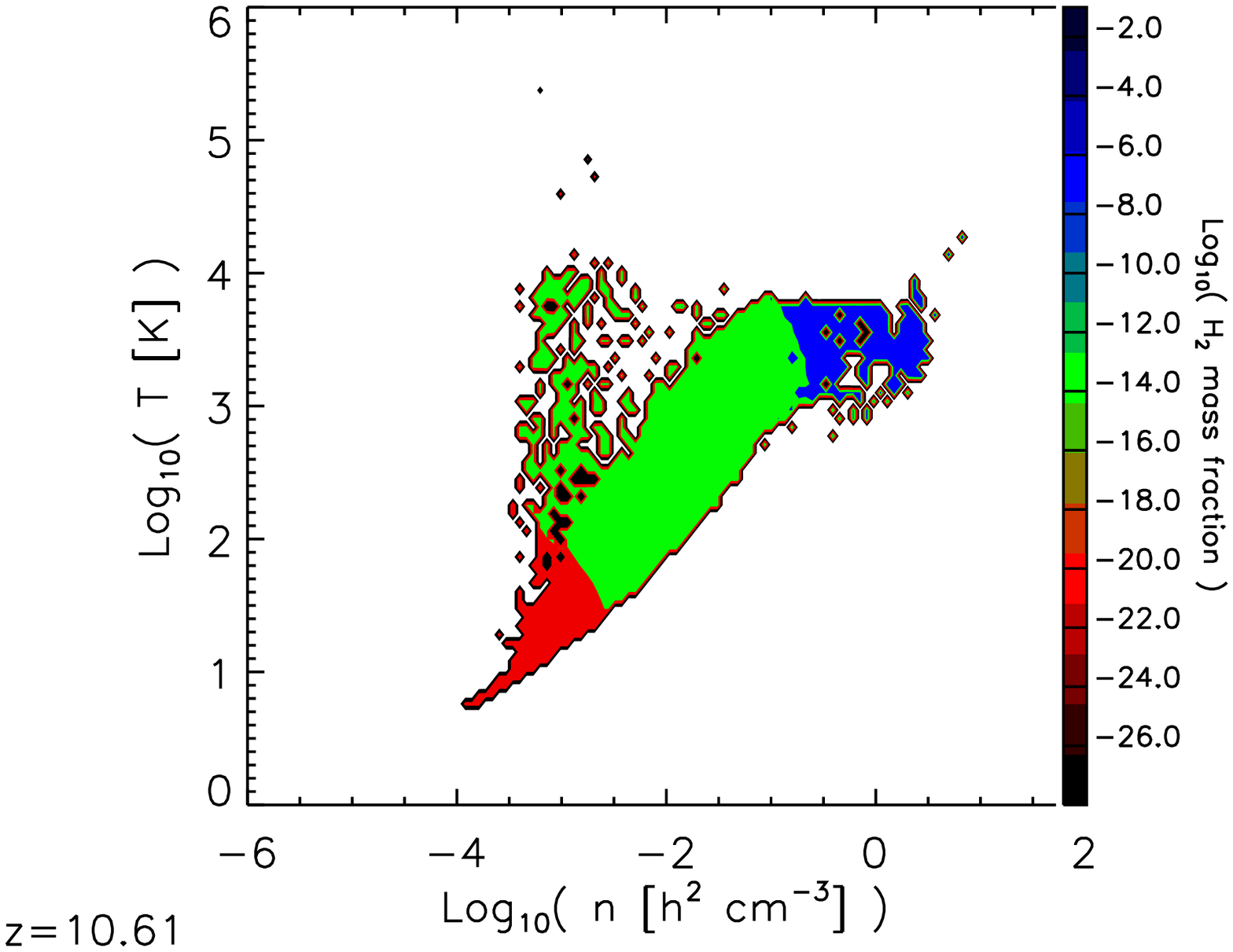}\\
\includegraphics[width=0.4\textwidth]{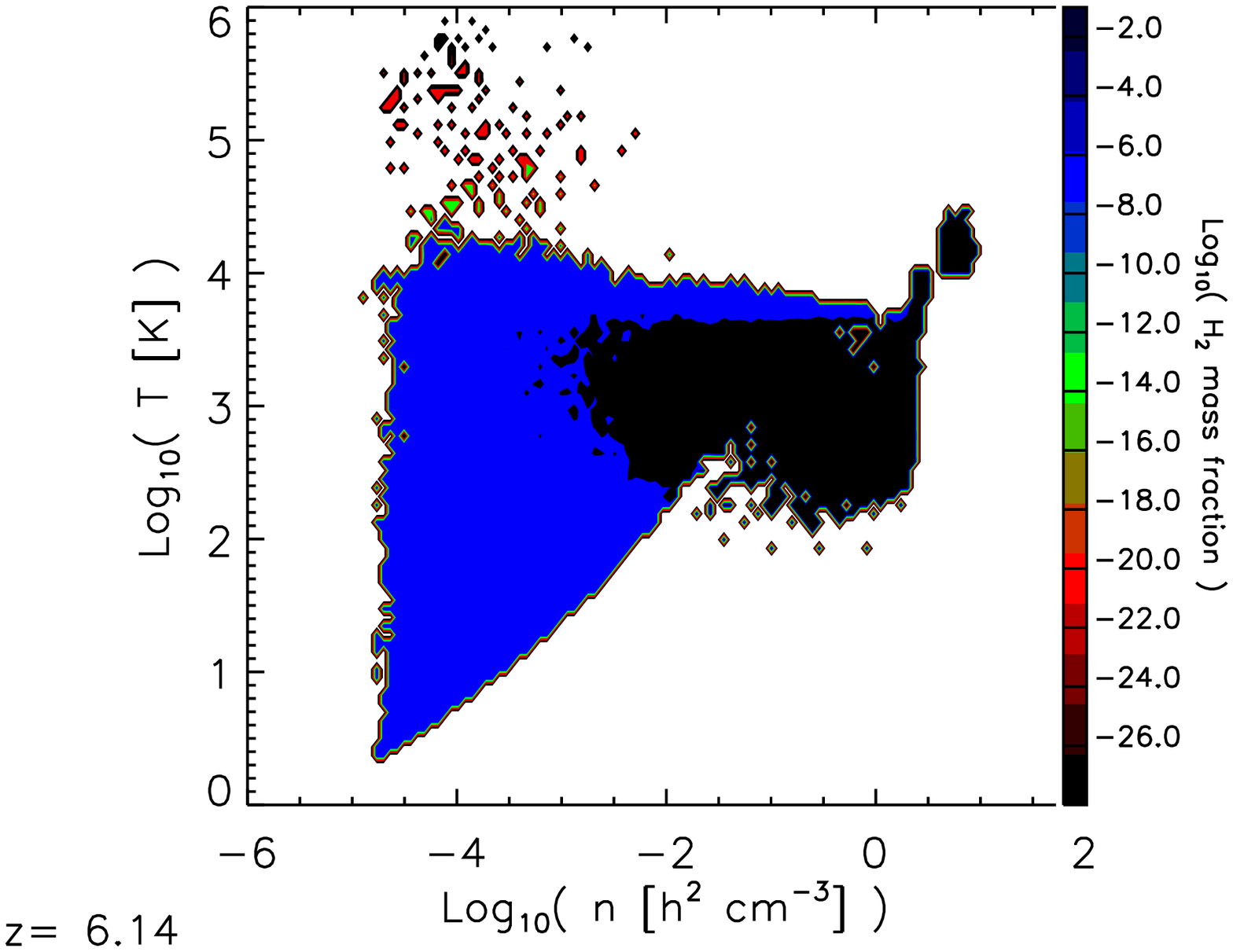}
\hspace{1.0cm}
\includegraphics[width=0.4\textwidth]{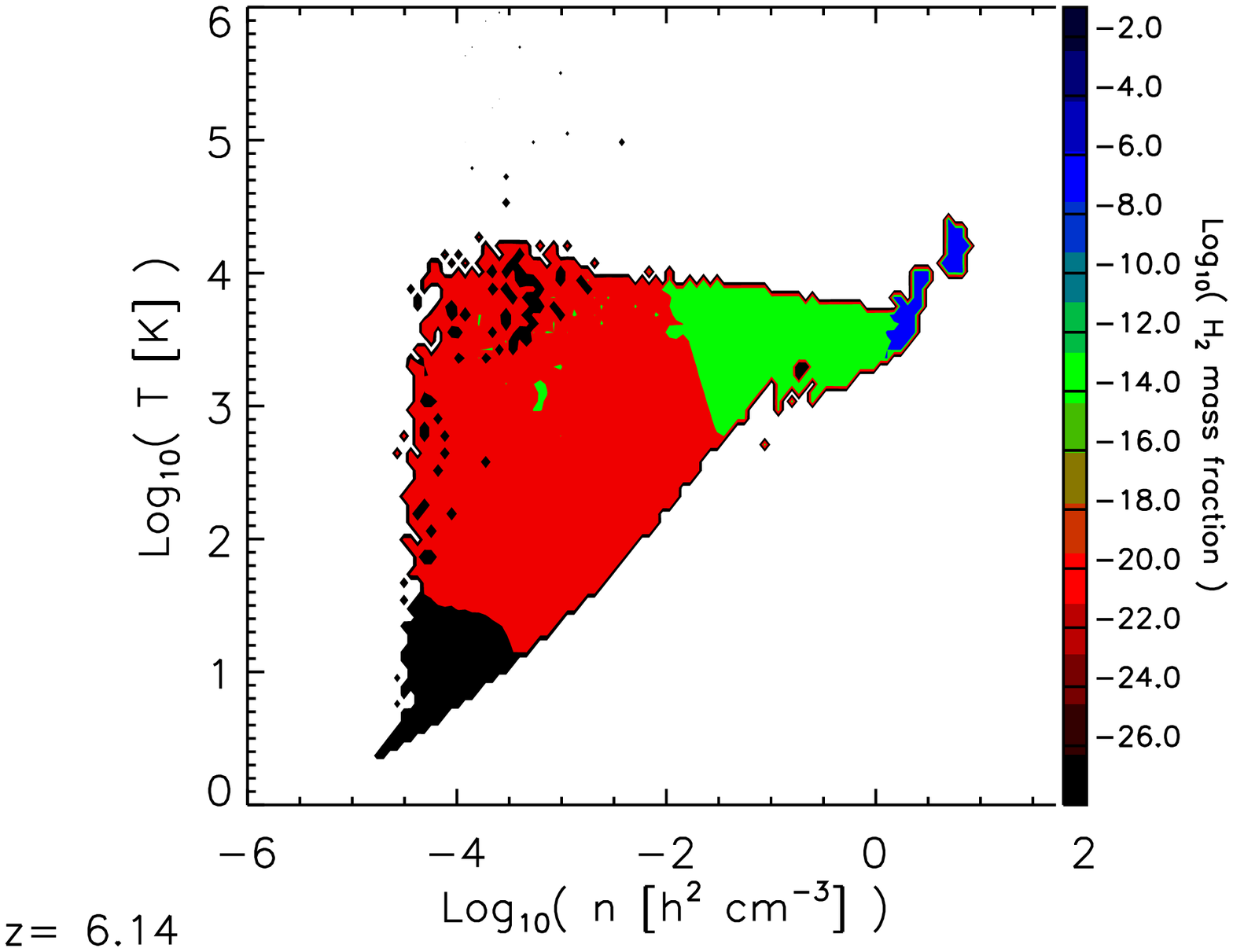}
\caption[]{\small Phase diagrams - temperature versus gas comoving number
  density at redshift $z \sim 10$ (top) and redshift $z \sim 6$
  (bottom) for a simulation without (left) and 
  with (right) ionizing radiation. The color shows the mass fraction
  of H$_2$. In the presence of ionizing radiation,  H$_2$ is depleted on
  all scales by many orders of magnitude. The effect is stronger at lower
  redshift due to the larger number of photons available.}
\label{fig:phase}
\efigs
\bfigs
\centering
\includegraphics[width=0.45\textwidth]{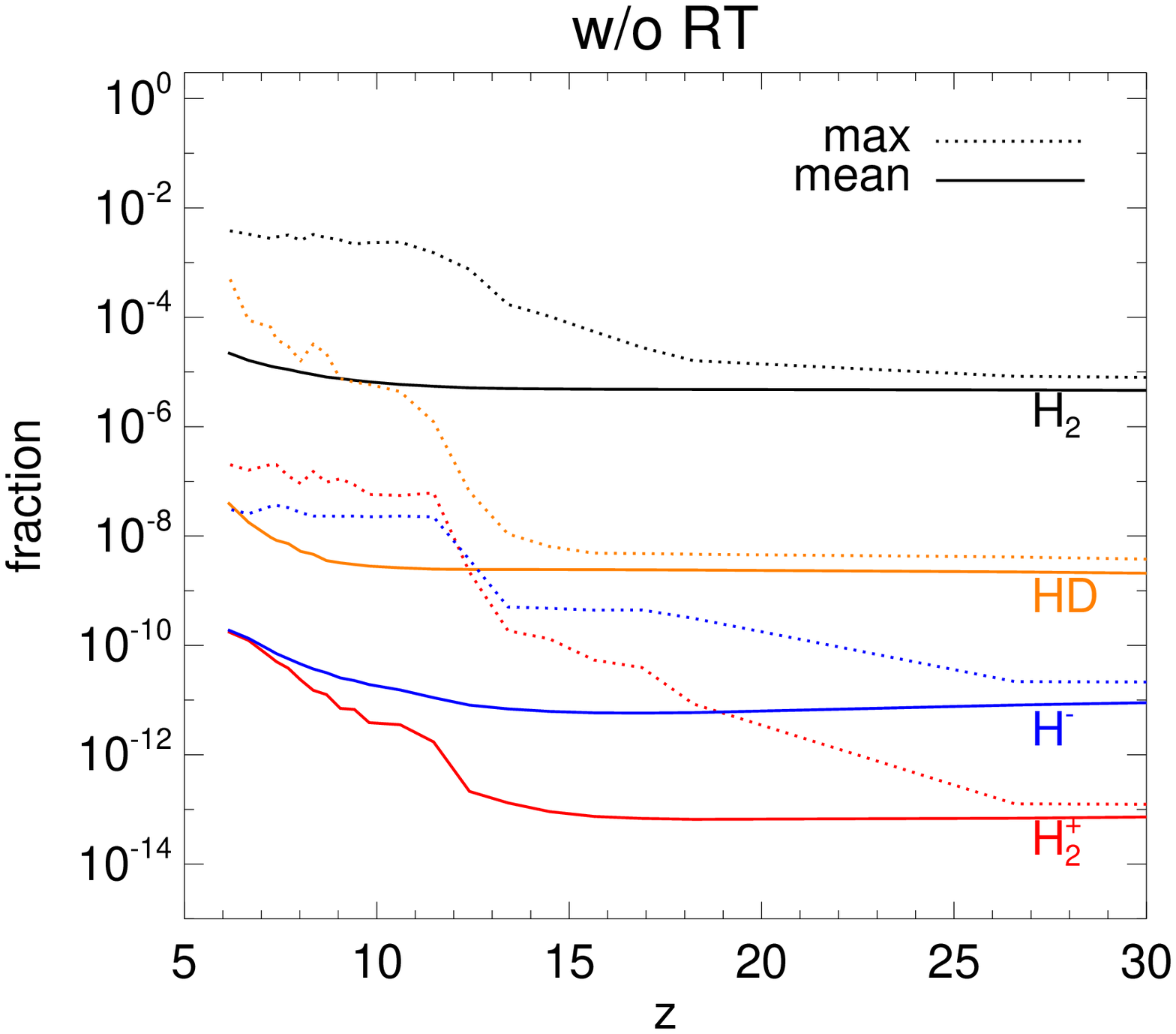}
\hspace{1.0cm}
\includegraphics[width=0.45\textwidth]{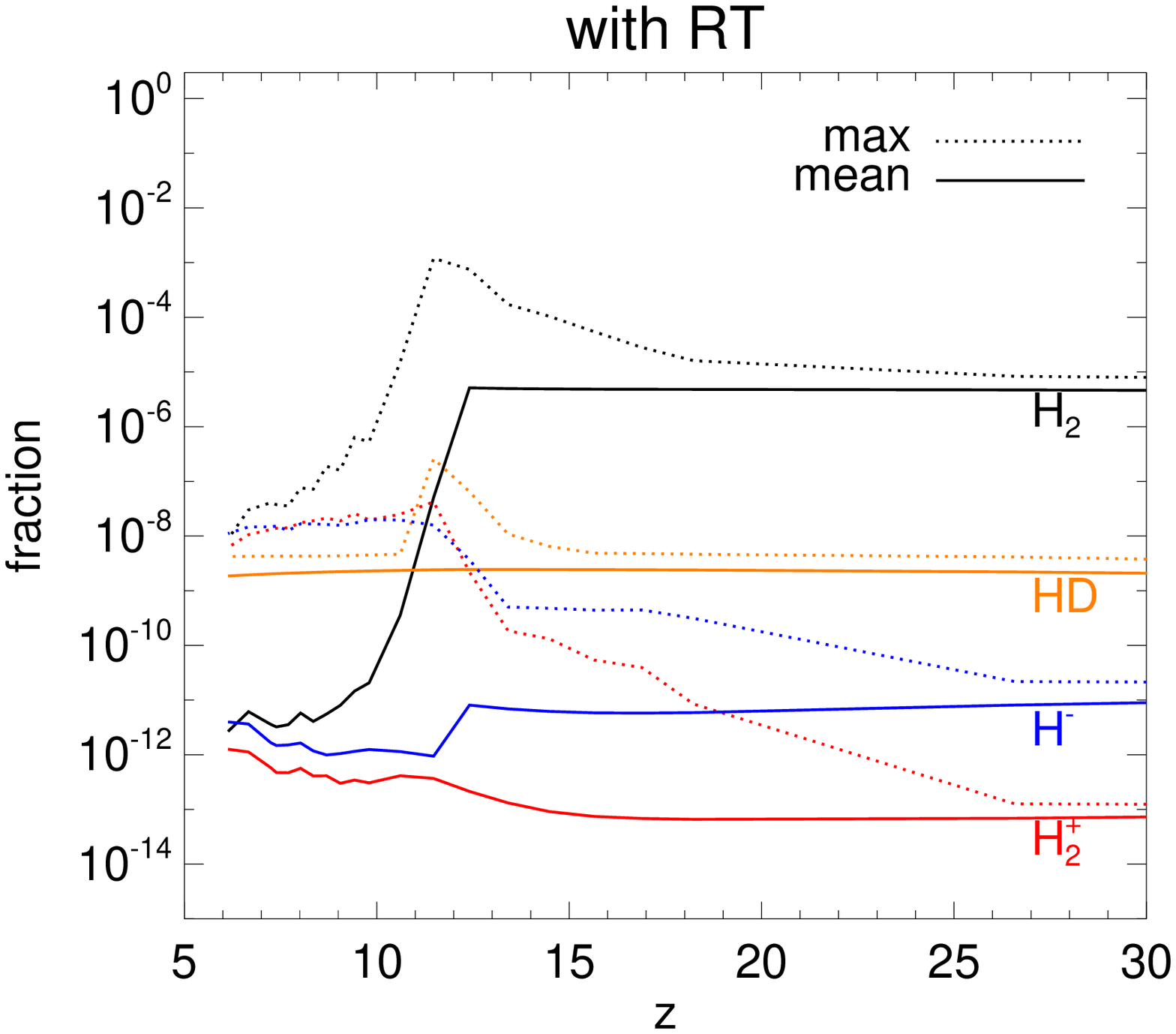}
\caption[]{\small Evolution of the mean and maximum H$_2$, HD, H$^-$ and
  H$_2^+$ fractions with redshift for the simulation with ionizing
  radiation (right) and without (left). When star formation sets in at
  redshift $z \sim 11$, the H$_2$ fractions drop fast, where both the
  mean and the maximum values are affected. The mean HD fraction
  stays constant in the presence of ionizing radiation and is
  increasing with lower redshift in the absence of radiation. The
  maximum follows the same trend, with the difference that it drops
  down as the first photos begin to propagate. There is only a small
  change in the evolution of the H$^-$ fraction. Finally, the H$_2^+$
  fraction is slightly suppressed from the ionizing radiation.}
\label{fig:comp_evolution}
\efigs

As a final application, we present results from a cosmological structure 
formation simulations in the framework of the standard $\Lambda$CDM model,
and we will also consider additional feedback mechanisms from star 
formation (as already mentioned in Sec.~\ref{Sect:introduction}).
We use a periodical comoving boxsize of $L_{\rm box} = 0.5 h^{-1} \rm
Mpc$ on a side with $2 \times 64^3$ gas and dark-matter particles 
(for a resulting spacial resolution of $\sim 0.4\,\rm kpc/{\it h}$ comoving),
sampled at the initial redshift $z=100$.
The runs include gravity, hydrodynamics, wind feedback, non-equilibrium
chemistry, and radiative transfer.
Stars are taken to be sources of ionizing radiation.
Since each star particle in the simulation represents a whole stellar
population with a Salpeter distribution, about 12{\%} of its mass is
in high-mass stars ($\gtrsim 10\msun$) that are able to produce UV photons.
The frequency distribution is assumed to be a black-body spectrum 
with an effective temperature of $3\times 10^4 \, \rm K$,
which corresponds to a  luminosity of approximately $8 \times 10^{48} \,
\rm photons \, s^{-1}$ per high-mass star in the stellar population
of the star particle.
We assume that gas particles are converted into stars once a critical density
of $\sim 10\,\rm cm^{-3}$ is reached, and the gas temperature is below $\sim 10^4\rm K$
to make sure that the gas is effectively cooling \cite[see details in][]{Maio2009}.
Star forming particles also experience SN-explosion feedback, 
which heats the gas above $\sim 10^5\,\rm K$, and wind feedback, 
which expels gas with a typical velocity of $\sim 500\,\rm km/s$
\cite[see also][and references therein]{Springel2005}.

A pictorial representation of the simulated box is given in Fig.~\ref{fig:maps},
where we show mass-weighted temperature slices through the simulation
volume at redshift $z=10.61$ and $z=6.14$ for the full run, 
including, in particular, both non-equilibrium chemistry and RT.
\\
The first sources are well visible close to the central part of the slice, 
and, while the structure growth proceeds, more sources are found in 
scattered places along the converging filaments.
Obviously, in a wider perspective (say on $\sim 100\,\rm Mpc$ scale), 
radiative sources would be much more uniformly distributed.
However, there are currently serious computational limitations for 
performing simulations with such large box sides and, simultaneously, with 
resolution good enough to resolve chemical evolution and radiative
transfer at the same time.
In the maps, the filamentary cold structures led by early molecular gas are 
well visible at temperatures around hundreds Kelvin.
In the densest regions, radiative effects from first stars heat the medium
above $\sim 10^4\,\rm K$ already by redshift $z\gtrsim 10$.
More and more star formation episodes appear at later stages and contribute
to the cosmic reionization process down to redshift $z\simeq 6$.
\\
Given the small size of our box, we can clearly focus on the
infalling phases of the cold material in the intersection of primordial
filamentary structures, and on the subsequent SN explosions, which heat the
gas and push material into the lower-density, void regions.
This helps understanding the different role played by the 
various feedback mechanisms, on the other side,
the lack of a very large box also leads to an insufficient number of stellar 
sources to fully complete reionization by $z\sim 6$.

In Fig.~\ref{fig:phase} we plot the phase diagrams (gas temperature
vs. number density) for the simulations with and without radiative
treatment.
The colors refer to the H$_2$ mass fraction. Molecular hydrogen
is dissociated in the presence of ionizing radiation (mostly in the 
Lyman-Werner band), and this effect increases with decreasing redshift,
due to the higher number of available photons.
As a consequence, this lowers the star formation process, since H$_2$ 
provides the largest part of the sub-$10^4\rm \, K$ cooling.
It is evident from the plots that without radiative feedback H$_2$ can
reach levels $\gtrsim 10^{-2}$ and boost early star formation.
In presence of radiation from stellar sources, molecular fractions decrease
of several orders of magnitude down to $\lesssim 10^{-4}$.
This effect is stronger for low-density gas, where molecules re-form more slowly.
Such behaviour is clear from the both cases shown, at redshift
$z=10.61$ and $z=6.14$.
In the latter case, a wider temperature spread, due to thermal heating of
the infalling gas at low density, appears as a consequence of the ongoing
structure formation.
Additionally, also SN-heated gas at $10^5-10^6\,\rm K$ is evident with
extremely low molecular fractions ($\lesssim 10^{-16}$).
\\
The lack of very cold gas in the more realistic case with both chemistry and RT 
suggests that low-temperature cooling by metals \cite[e.g.][]{Maio2007} could be 
an important mechanism to sustain star formation after the first generation of 
stars \cite[][]{Maio2010}\footnote{
  One could also rely on reionized gas \cite[][]{Yoshida2007} to form subsequent
  baryonic structures.
}.
To better underline the impacts of radiative effects on early chemistry, we 
also show in Fig.~\ref{fig:comp_evolution} the redshift evolution of the
mean and maximum number fractions of the two main molecules, H$_2$ and HD, 
and of the basic species for the different channels of molecular formation,
H$^-$ and H$_2^+$.

As mentioned previously, after early star formation sets in, around redshift
$z \sim 12$, the H$_2$ fractions drops by several orders of magnitude (from
average values of $\sim 10^{-5}$ down to $\lesssim 10^{-10}$), as
ionizing radiation starts to propagate in the simulated volume.
This is not seen in the case without radiation, where H$_2$ increases
almost monotonically, with peak values of $\sim 10^{-2}$ at $z\simeq 6$.

The mean HD fraction stays roughly constant in the presence of ionizing 
radiation, but it is increasing in the case without radiation. 
When comparing peak values, one sees that, at $z\simeq 12$, HD has fractions
of $\sim 10^{-6}$ in both cases, but in the RT case, it is significantly destroyed
when photons begin to propagate. Instead in the non-RT case its maximum values 
catch up with H$_2$.

The H$^-$ fraction is crucial for H$_2$ formation via the H$^-$-channel:
in presence of RT its mean values are lowered down to $\sim 10^{-12}$,
so they cannot catalyze efficiently further molecule formation, while
in the non-RT case, H$^-$ would increase by two orders of magnitude.

Similarly, H$_2^+$ catalyzes H$_2$ formation via the H$_2^+$-channel, but
photon propagation destroys molecules and inhibits H$_2^+$ formation.
As a comparison, in the non-RT run H$_2^+$ reaches mean values of $\sim 10^{-10}$,
a couple of orders of magnitude larger than in the RT run. Maximum
fractions for H$^-$ and H$_2^+$ are similarly suppressed of about one order of 
magnitude.

As a conclusion, the radiative feedback is responsible for molecule
dissociation. Molecules are easily depleted not only from the external
sources, but also from the central sources which have just been formed.
This is justified by the fact that catastrophic molecular cooling sets in 
already at densities $\gtrsim 1\,\rm cm^{-3}$, when the material is  
optically thin, and thus there is no significant gas shielding
preventing H$_2$ and HD dissociation.
In fact, in the simulations presented here, radiative feedback 
becomes effective at densities $> 10\,\rm cm^{-3}$.
In the innermost central regions (i.e. below a few hundreds of comoving
parsec, not sampled by our simulations because of resolution limits),
densities could rise up to values greater than $\sim 10^8\,\rm cm^{-3}$,
become optically thick, and produce significant shielding\footnote{
If we consider that primordial haloes have radii of a few kpc, the fraction 
of volume interesting for such events is very small:
for a protogalaxy having a radius of $\sim\,\rm kpc$ and developing a dense 
core within $\sim 100\,\rm pc$, the volume fraction of high-density, possibly
shielded gas is $\lesssim 10^{-3}$.
}.
This should not affect the overall molecular destruction over $\sim$~kpc scales 
by external sources, though.
\\
We stress that precise, quantitative assessments about the net effects of 
radiative feedback on the surrounding gas are difficult, because one should
be able to prove RT on very different scales within N-body/hydro chemistry 
calculations.
While destruction of molecules in low-density intergalactic medium (IGM) is 
intuitive and expected, the behaviour in the innermost cores of collapsed 
objects is quite debated and strongly dependent on the local density regime.
In particular, as pointed out by \cite{Susa2007}, the ignition timing of 
the source star is crucial in collapsing clouds:
if the ignition takes place when the central density is larger than 
$\sim  10^3-10^4 \,\rm cm^{-3}$, the collapse cannot be halted by radiation.
However, following in great detail the stellar-core formation and the 
ignition of nuclear reactions which lead the birth of the star is nowadays 
still prohibitive.\\
As a consequence, cosmological simulations are inevitably affected by
resolution issues, mostly at large densities, that limit the possibility of 
fully understanding how the various physical processes interplay in the 
very final stages of star formation.
This also affects our capabilities of quantifying photon escape from the
dense star forming regions.
It is reasonable to think that in realistic conditions the inner cores will
be optically thick and only after the photons will have heated up the cloud
and thermalized with it they will be able to escape and dissociate molecules
in the diffuse IGM.
Definitive answers on this topics are still lacking, though.
Furthermore, these many uncertainties even lead to differences of several 
orders of magnitudes when comparing mass estimates of haloes affected by 
radiation \cite[e.g.][]{Ciardi2000b,Kitayama2001,Dijkstra2004}.

%************************************************************************

\section{Summary and conclusions}\label{Sect:discussion}

%************************************************************************

Cosmic gas and star forming processes are fundamental keywords in modern 
Astrophysics. They are, however, very complicated to study because
they lie in the highly non-linear regime.
Thus, simple perturbative approaches do not give relevant information about gas 
collapse and baryonic-structure growth.
To understand them better, it is crucial to consider
all the involved physical and chemical mechanisms.
In case of early structure formation, primordial chemistry has to be taken into
account, since it is mainly via H-derived molecules that first objects can 
aggregate gas and form stars.
Furthermore, it is also necessary to self-consistently include the RT
of the photons produced by stellar sources.
Indeed, these travel into the cosmic medium and can ionize the surrounding gas
or dissociate formed molecules.
As the impacts of RT on the chemical evolution of the early Universe are expected 
to be significant, it is important to couple such calculations in order to get
a reliable picture.
\\
In this work, we have presented numerical methods of coupled chemistry treatment
and multi-frequency RT, applied to N-body, hydrodynamical simulations.
We used the SPH code {\small GADGET3} -- an extended version of the
publicly available code {\small GADGET2} \cite[][]{Springel2005} --
including the RT implementation of \cite{Petkova2009}, and
the non-equilibrium chemistry implementation of \cite{Maio2007}, following
e$^-$, H, H$^+$, H$^-$, He, He$^+$, He$^{++}$, H$_2$, H$_2^+$, D, D$^+$, HD, HeH$^+$.
\\
After a detailed description of the coupling between the RT treatment 
(Sect.~\ref{Sect:radtrans}) and the non-equilibrium chemical treatment 
(Sect.~\ref{Sect:chemistry}), we have performed convergency tests 
(Sect.~\ref{Sect:simulations}) and cosmological applications 
(Sect.~\ref{Sect:cosmo}) of our code.

We have started (in Sect.~\ref{Sect:ss}) with the expansion of an ionized sphere
around a stellar-type source emitting at a temperature of 
$\sim 3\times 10^4\,\rm K$ in a uniform-density gas.
We have traced the element and molecule evolution with time and compared to
analytical results
(in Fig.~\ref{fig:SST_um_evol} and Fig.~\ref{fig:SST_um}).
We found that analytical analyses based on H-only gas at a constant temperature give 
a sufficient criterion to predict the evolution of the I-front.
The different species reach different ionization radii:
we use the Str{\"o}mgren radius defined for H-only gas at $10^4\,\rm K$,
as a reference.
Deuterium species recombine at a radius comparable (within 10 per cent) 
to the Str{\"o}mgren radius, while He gets completely neutral at $\sim 2/5$
the Str{\"o}mgren radius.
H$_2$ has an ionization radius that is typically larger than the Str{\"o}mgren radius 
of about $40$ per cent ($\sim 7\,\rm kpc$ vs $\sim 5\,\rm kpc$).
The compression of gas due to the propagating I-front enhances molecular species
(H$_2$ and HD) in the outer layers, where temperatures are around $\sim 10^3-10^4\,\rm K$.
The high values of the creation rates in these temperature regimes make H$_2$ increase 
of a few orders of magnitude and HD, as well.

The second test (see Sect.~\ref{Sect:abundances}) is a mean-density 
cosmological evolution, where the CMB was assumed to be the only source of 
ionizing radiation.
We found that (Fig.~\ref{fig:cosmo_evolution}), because of the low cosmic background
emission, the CMB does not affect significantly the abundance evolution, even when
considering photon propagation at very-high redshift.

As applications of the numerical methods previously described, we have studied
(Sect.~\ref{Sect:cosmo}) the effects of RT on chemical evolution of primordial
gas (e.g. Fig.~\ref{fig:maps}).
We performed cosmological simulations of structure formation with and without ionizing 
radiation from stellar sources, and checked the consequences for molecule formation 
and destruction.

We found that the presence of ionizing radiation from stars depletes molecular
hydrogen up to several orders of magnitude (see Fig.~\ref{fig:phase}), by inhibiting
the main formation paths (the H$^-$-channel and the H$_2^+$-channel) and by dissociating
it via Lyman-Werner radiation.
Our results are consistent with other studies, e.g.
\cite{Wise2007,Johnson2007,Ahn2009,TrentiStiavelli2009,Whalen2010,Latif2011},
who studied the impacts of the UV background in destroying down the H$_2$ molecule.
In addition, we found that also other molecules formed in pristine gas, like e.g. HD,
are strongly suppressed by radiative feedback (Fig.~\ref{fig:comp_evolution}).

There are large discrepancies in the quantitative assessments of the impacts of 
radiative feedback on baryonic structure formation in the current literature.
In fact, different works \cite[e.g.][]{Ciardi2000b,Dijkstra2004}
show uncertainties of several orders of magnitudes on the basic
estimates of the masses of the haloes affected by radiation.
Such determinations could even be highly biased by their different
``post-processing'' approaches.
Indeed, in order to have large statistics or to circumvent numerical complications,
these studies do not consider hydrodynamics self-consistently coupled with RT and
non-equilibrium chemistry.
That could be problematic for getting to reliable conclusions, since, as shown by 
our numerical simulations (Sect.~\ref{Sect:cosmo}), RT effects on gas evolution could
be quite important and affect star forming regions in a non-negligible way.
\\
The destruction of early molecules by stellar sources has a deep impact
on the following star formation processes, because it hinders the successive
birth of metal-free stars.
This also implies the need of different viable low-temperature coolants, like metals
\cite[e.g.][]{Maio2007}, or the presence of reionized gas \cite[][]{Yoshida2007} 
to sustain star formation at later times.
The obvious expectation is, then, that the popIII star formation rate will drop
\cite[even more heavily than expected by][]{Maio2010,Maio2011b,Hasegawa2009,Tornatore2007,OShea2008,Johnson2011}.
However, further and more detailed investigations of high-resolution, numerical
simulations are required to draw final and definitive conclusions on this topic.

%************************************************************************

\section*{acknowledgments}
We acknowledge useful discussions with 
Benedetta Ciardi,
Mark Dijkstra,
Klaus Dolag,
Edoardo Tescari,
Dan Whalen.\\
The analysis and the simulations have been performed on the MPA
AMD Opteron machines at the Garching Computing Center
(Rechenzentrum Garching, RZG).
This research has made use of NASA's Astrophysics Data System and of
the JSTOR Archive.

%************************************************************************

\bibliographystyle{mn2e}
\bibliography{bibl,bibl2}

\label{lastpage}
\end{document}